\documentclass[10pt]{article}
\usepackage{upgreek}
\usepackage[utf8]{inputenc}
\usepackage[english]{babel}
\usepackage{graphicx,amsmath,amsfonts,amssymb,dcolumn,amsthm,slashed,bbm,xspace,pgffor,tikz,mathbbol,latexsym,cite,braket}

\usepackage{microtype,tikz}
\usepackage{lmodern}
\usepackage[normalem]{ulem}
\usepackage{xcolor}
\usepackage[colorlinks=true,citecolor=blue,urlcolor=blue,linkcolor=blue]{hyperref}
\usepackage[hang]{footmisc} %removes the footer margin space
\usepackage[affil-it]{authblk}
\usepackage[compat=1.1.0]{tikz-feynhand}
\usepackage{multicol}
\usepackage{microtype,diagbox}

\newcommand{\orcidicon}{%
	\begin{tikzpicture}
	\draw[lime, fill=lime] (0,0)
		circle [radius=0.16]
		node[white] {{\fontfamily{qag}\selectfont \tiny ID}};
	\draw[white, fill=white] (-0.0625,0.095)
		circle [radius=0.007];
	\end{tikzpicture}	\hspace{-2mm}
}

\newcommand\orcidroldao{{\href{https://orcid.org/0000-0003-3978-532X}{\orcidicon}}}
\newcommand\orcido{{\href{https://orcid.org/0000-0002-2685-3349}{\orcidicon}}}

\numberwithin{equation}{section}

\usepackage{float}
\begin{document}

\title{\textbf{Phase transitions at high and low densities for a rotating QCD matter from holography}}

\author[1,2]{Octavio C. Junqueira\protect\orcido\thanks{\href{mailto:octavioj@pos.if.ufrj.br}{octavioj@pos.if.ufrj.br}}}
\author[1]{Roldao da Rocha\protect\orcidroldao\thanks{\href{mailto:roldao.rocha@ufabc.edu.br}{roldao.rocha@ufabc.edu.br}}}
\affil[1]{Center of Mathematics, Federal University of ABC, 09210-580\\ Santo Andr\'e, S\~ao Paulo, Brazil}
\affil[2]{Center of Physics, Federal University of ABC, 09210-580\\ Santo Andr\'e, S\~ao Paulo, Brazil}

\date{}
\maketitle

\begin{abstract}

We applied the exact Andreev soft-wall holographic model to investigate phase transitions in rotating strongly interacting matter at high and low densities. Using the dual description of hadronic matter and quark-gluon plasma via thermal and charged black holes in five-dimensional AdS space with cylindrical symmetry, we find that for relativistic rotations exceeding 16\% of the speed of light, crossover transitions emerge in the low-density regime up to a critical baryon chemical potential $\mu_{CPB}$. These smooth transitions, governed by the negative QCD $\beta$-function, describe a mixed phase of confined and deconfined matter with different angular momenta evolving into a pure plasma at very high temperatures. For $\mu \geq \mu_{CPB}$, first-order transitions dominate, following the critical-temperature curve of non-rotating matter. The critical point separating the low-density crossovers from high-density first-order transitions is numerically estimated as $(\mu_{CPB}, T_{CP}) = (363.554, 58.507)\,\text{MeV}$.

\end{abstract}

\section{Introduction}
\label{sec1}

The study of strongly interacting matter under extreme conditions is a central challenge in modern high-energy physics. Heavy-ion collision experiments at RHIC, the LHC, and future facilities such as FAIR and NICA indicate that QCD matter undergoes a transition from a confined hadronic phase to a deconfined quark–gluon plasma (QGP) \cite{Critelli:2017oub,Shuryak:2008eq,Busza:2018rrf}. The characteristics of this transition depend sensitively on both temperature and baryon density, and the existence and precise location of the QCD critical point remain open questions \cite{Fukushima:2010bq,Toniato:2025gts,Bzdak:2019pkr}. While lattice QCD establishes that the transition at zero baryon chemical potential is a smooth crossover \cite{Aoki:2006we,Borsanyi:2010bp,Cucchieri:2007rg,Torrieri:2007qy}, finite-density calculations are limited by the sign problem \cite{Muroya:2003qs}, making effective and nonperturbative models essential for exploring the QCD phase diagram \cite{Krein:2017usp}.

Holographic approaches provide a natural framework to address this challenge. The AdS/QCD correspondence, derived from the AdS/CFT principles \cite{Maldacena:1997re,Gubser:1998bc,Witten:1998qj}, maps strongly coupled, nonperturbative QCD at the boundary to weakly coupled gravity in a five-dimensional AdS$_5$ bulk \cite{Rougemont:2017tlu,Branz:2010ub,daRocha:2024lev,Colangelo:2008us,Brodsky:2014yha,Rougemont:2023gfz}. Bottom-up models, particularly the soft-wall AdS/QCD model, efficiently incorporate nonperturbative QCD phenomena such as confinement, spontaneous chiral symmetry breaking, and hadronic Regge trajectories \cite{Karch:2006pv,Gursoy:2007cb,Bartz:2018nzn,daRocha:2021xwq,Ballon-Bayona:2023zal,Aref'eva2018May,Csaki,Gherghetta:2009ac,Erlich:2005qh,Sakai2005Apr}. Although QCD lacks exact conformal symmetry, soft-wall constructions keep up with chiral symmetry breaking and the QCD scale with suitable choices of dilaton profiles and warp factors, allowing parameters to match phenomenological data \cite{Bernardini:2018uuy,Braga:2020opg,Ferreira:2019inu,Shukla:2023pbp,Ferreira:2020iry}. Finite-density extensions often employ the AdS–Reissner–Nordström approximation \cite{Lee:2009bya,Colangelo:2010pe}, but exact solutions such as Andreev’s charged rotating black hole (BH) in AdS$_5$ provide a more accurate holographic description \cite{Andreev:2010bv}. 
The fluid/gravity correspondence further connects holographic QCD to dissipative relativistic hydrodynamics that describe the QGP \cite{Bemfica:2020xym,Rocha:2022ind,Kovtun:2019hdm,Bhattacharyya:2007vjd,Policastro:2002se}. 

In ultrarelativistic, noncentral heavy-ion collisions, the system acquires substantial angular momentum \cite{STAR:2017ckg}, generating vorticity in the QGP. This rotation modifies the initial longitudinal velocity profile, enhances elliptic flow, and affects the expansion dynamics by interacting with shear and bulk viscosities \cite{Abboud:2023hos}. Observable consequences include global polarization of hadrons and the chiral vortical effect \cite{Jiang:2016woz,Becattini:2007sr,Kharzeev:2015znc,GoncalvesdaSilva:2017bvk}. AdS/QCD studies have provided crucial insights into these rotational effects \cite{Braga:2022yfe,Braga:2023qee}, making rotation an essential ingredient for realistic modeling of strongly coupled QCD matter. 
Rotation can also reshape the phase structure of QCD matter in striking ways. The angular velocity suppresses the chiral condensate and may generate a critical point in the $(T,\omega)$ plane, while mixed inhomogeneous phases have been proposed in rotating systems \cite{Fujimoto:2021xix,Zhao:2022uxc}. Hadron resonance gas results indicate a decrease of the deconfinement temperature with rotation \cite{Chen:2017xrj}. Holographic QCD also reveals modifications in deconfinement patterns, providing relations between the critical temperature and the angular velocity. Recent lattice QCD simulations also corroborate the modification of the deconfinement temperature in the QGP under rotation  \cite{Braguta:2023iyx}.

Gauge/gravity duality also predicts phase transitions that relate the QGP to the hadronic phase \cite{Noronha:2009ud}. In holographic QCD, the confinement/deconfinement transition is interpreted as a Hawking–Page transition between thermal AdS and an AdS BH \cite{Herzog:2006ra}. The soft-wall model captures linear Regge trajectories, confinement, and other infrared QCD features \cite{DaRold:2005mxj,Li:2013oda}. Rotation introduces modified gauge-field dynamics that qualitatively alter BH thermodynamics and the boundary phase structure. 
In this work, we numerically analyze Hawking–Page transitions in the exact Andreev soft-wall model at finite density and rotation. Using the full on-shell action, including gravitational, dilaton, and Abelian gauge-field contributions, we compute the renormalized action density difference between rotating charged AdS BHs and thermal AdS. This identifies the dominant thermal saddle and maps the QCD-like phase diagram in the $(T,\mu)$ plane for various angular velocities $\omega$. We find three distinct regimes: at zero density, a Herzog-type first-order transition persists with rotation affecting only the critical temperature via relativistic redshift; at high density, the first-order transition survives, but the critical temperature decreases with rotation and the maximal density for Hawking–Page transitions becomes rotation-dependent; at low density and high angular velocity, $\omega l \gtrsim 0.16$, Hawking–Page transitions disappear over a finite chemical potential range, producing smooth crossovers. The interplay of these regimes generates a holographic critical point, whose location is estimated after calibrating the holographic energy scale.

The paper is organized as follows: Sec. \ref{sec2} contains an overview of the construction of the regularized rotating charged BH action density in AdS space, with its Hawking temperature obtained from the surface gravity formula. In Sections \ref{sec3}, \ref{sec4}, and \ref{sec5}, we describe the first-order transitions that occur for non-rotating matter, at large density and at zero density, respectively. In Sec. \ref{sec6}, we analyze  the Gibbs free energy, entropy and specific heat from holographic renormalization, besides smooth transitions (crossovers) in the low-density regime, accounting for relativistic rotations. The non-relativistic limit of the phase transitions at low and high densities is also studied. In Sec. \ref{sec8} we obtain a numerical estimate for the critical point and pave the comparison with lattice QCD, effective models, and holography in Sec. \ref{sec8}. Finally, Sec. \ref{sec9} contains our conclusions. 

\section{Rotating charged BH in the exact Andreev's soft wall model}
\label{sec2}
The gravitational dual of a rotating QGP at finite density is given by a charged BH with nonzero angular momentum in five-dimensional AdS spacetime. This space is an exact solution to Einstein's field equations with a negative cosmological constant $\Uplambda = - \frac{12}{L^2}$ and a constant Ricci scalar $R = - \frac{20}{L^2}$, where $L$ denotes the curvature radius of AdS. Assuming cylindrical symmetry, for matter rotating with a uniform angular velocity $\omega$ around a hypercylinder with radius $l$, the system can be described by the following charged BH metric in the canonical form
\cite{Zhou:2021sdy, LEMOS199546, BravoGaete:2017dso}:
\begin{eqnarray}\label{canon}
	ds^2 = N(z,q)\, dt^2 + \frac{L^2}{z^2}\frac{dz^2}{f(z,q)} + R(z,q)\,\left(d\phi \!+\! P(z,q) dt\right)^2 + \frac{L^2}{z^2} \sum_{i\,=\,1}^2 dx_i^2\;,
\end{eqnarray}
with 
\begin{eqnarray}
	N(z,q) &=& \frac{L^2}{z^2} \frac{(1-\omega^2 l^2) f(z,q)}{1- \omega^2 l^2 f(z,q)}\;, \\
	R(z,q) &=& \frac{L^2\,l^2\,\gamma^2}{z^2}\left(1  -  f(z,q) \omega^2 l^2\right)\;, \\
	P(z,q) &=& \frac{\omega(1-f(z,q))}{1- \omega^2 l^2f(z,q)}\;,
\end{eqnarray}
where $\gamma = 1/\sqrt{1 - l^2 \omega^2 }$ stands for the Lorentz factor, and 
\begin{equation}\label{f(z)}
f(z,q) = 1 - \frac{z^4}{z_h^4}-q^2\, z_h^2 z^4 + q^2 z^6\;,
\end{equation}
being $z_h$ the location of the BH event horizon, such that $f(z_h,q) = 0$, whereas the $q$ parameter encodes the BH charge. On the other hand, the gauge dual of the hadronic phase is the thermal AdS spacetime, described by the metric \eqref{canon}, taking the limit $f(z,q) \rightarrow 1$.

Defining $N(z)=-h_{00}(z)$ \cite{Zhou:2021sdy}, the Hawking temperature of the rotating charged BH can be obtained from the surface gravity formula,
\begin{eqnarray}\label{HTrot}
	T(q, \omega) &=&  \frac{\left\vert\kappa_G\right\vert}{2\pi}  = \frac{1}{4\pi}\lim_{z\rightarrow z_h} \left\vert  \sqrt{\frac{g^{zz}}{-h_{00}(z)}}\partial_z h_{00}\right \vert = \frac{1}{\pi z_h} \left(1-\frac{{q}^2 z_h^6}{2}\right) \sqrt{1-\omega^2 l^2}\;, 
    \end{eqnarray}
where $\kappa_G$ is the surface gravity, and $g^{zz}$ denotes the bulk component of the cylindrical BH inverse metric \eqref{canon}. It emulates the temperature of the hydrodynamic relativistic fluid flow in thermal equilibrium, describing the QGP.  The condition for the temperature to be positive, thus, requires the upper bound  
\begin{equation}\label{positivity}
  z_h \leq \left(\sqrt{2}/q\right)^{1/3}\;.
\end{equation}
For a compactified time coordinate, the BH time period is given by $\beta = 1/T$, being $T$ the BH temperature \eqref{HTrot}.  If one  requires that the asymptotic limits of both the thermal AdS and the BH AdS geometry in the rotating system equal each other at $z=\epsilon$, with $ \epsilon \to 0$, then the thermal AdS period reads  
\begin{equation}\label{betaAdS}
\beta_{AdS} (q,\omega) = \beta(q, \omega) \sqrt{f(\epsilon, q)}\;,
\end{equation}
which defines the Hawking-Page (HP) transitions between the BH and thermal AdS geometries, according to the action densities of each space  \cite{Barbosa-Cendejas:2018mng}.

In the holographic soft wall AdS/QCD model \cite{Karch:2006pv}, the five-dimensional  gravitational action in Euclidean space can be written as 
\cite{Herzog:2006ra,BallonBayona:2007vp}
\begin{equation}\label{action1}
	I_G = - \frac{1}{ 2 \upkappa^2} \int_0^{z_h} dz\int d^4x \sqrt{g} e^{-\Phi}\left( R - \Uplambda \right) \;,
\end{equation}
where $\Phi(z) = cz^2$ denotes the dilaton field breaking conformal symmetry and introducing the IR mass scale $\sqrt{c}$. Also, $\upkappa$ stands for the gravitational coupling associated with Newton's gravitational constant. The determinant of the metric, for both AdS spacetimes, is given by $g = l^2 L^{10}/z^{10}$. Taking into account the expression relating the AdS curvature and the cosmological constant, the on-shell gravitational  action reads 
\begin{equation}\label{action1onshell}
	I_{G_{\textsc{on-shell}}} = \frac{4l^2L^3}{\upkappa^2} V_{3D} \int_0^{\beta_s} dt \int_0^{z_h} dz\, \frac{e^{-cz^2}}{z^5}\;\,,
\end{equation}  
where $\beta_s$ denotes the period associated with the corresponding space, and $V_{3D}$ is the spatial bulk volume. The thermal AdS geometry has no event horizon, therefore $z_h \rightarrow \infty$ in this space.

In a system comprising quarks, the gauge vector field $V_\mu$ living in AdS space can be introduced to account for the BH charge. The five-dimensional action governing these  gauge vector fields is given by \cite{Braga:2015jca}
\begin{equation}\label{actionVF}
    I_{\textsc{VF}} = -\frac{1}{4g_5^2}\int_0^{z_h} dz \int d^4x \sqrt{g} e^{-\Phi} F_{MN}F^{MN}\;,
\end{equation}
where the gauge field strength is given by $F_{MN} = \partial_{M} A_N - \partial_N A_M$. The time component of $A_\mu$ works as the source of the correlation functions of the gauge theory density operator. This way, $A_0$ is interpreted as the quark chemical potential ($\mu$) correlated to the quark density $J^0=\bar{\psi}^\mu \gamma^0 \psi^\mu$ in the bulk. The  exact Andreev's solution of the gauge field equation of motion for the total $I = I_G + I_{\textsc{VF}}$ with the metric \eqref{canon} reads \cite{Andreev:2010bv, Junqueira:2025xnx}
\begin{eqnarray}
    A_0 &=& \gamma(\omega l) A_0^{\textsc{nrs}} \;,\label{A0}\\
    A_\phi & = &-l^2\omega \gamma(\omega l) A_0^{\textsc{nrs}}\label{Aphi}\\
    A_{x_1} &=& A_{x_2} = A_z = 0\;, \label{AiAz}
\end{eqnarray}
where $A^{\textsc{nrs}}_0$ is the time component of the  gauge vector field for the non-rotating system, namely, 
\begin{eqnarray}\label{A0final}
    A_0^{\textsc{nrs}}(z) = i\mu \left( \frac{e^{c z_h^2} - e^{c z^2}}{e^{c z_h^2}-1}\right)\;,
\end{eqnarray}
where $\mu = A_0^{\textsc{nrs}}(0)$ is the quark chemical potential. The Dirichlet boundary condition $A_0(z_h) = 0$ is satisfied, and is consistent with a gauge field with regular norm  
\cite{Horigome:2006xu, Nakamura:2006xk, Hawking:1995ap, Ballon-Bayona:2020xls}. This boundary condition defines the relation between the $q$ parameter and the chemical potential, as:
\begin{eqnarray}\label{qmu}
    \frac{\eta q}{c} = \frac{\mu}{e^{c z_h^2}-1}\;,  
\end{eqnarray}
with $\eta = \sqrt{\frac{3 g^2_5 L^2}{2 \upkappa^2}}$, which relates the BH charge $Q = \eta q$ with the quark chemical potential, often used to describe the QCD phase diagram in holographic models  \cite{Lee:2009bya}. 

The gauge-invariant quark chemical potential in the rotating system can be defined by the expression  \cite{Chen:2020ath,Zhao:2022uxc} 
\begin{equation}
    \mu^\prime = \lim_{z \to 0}A_\mu \chi^\mu - \lim_{z \to z_h}A_\mu \chi^\mu,
\end{equation}    
where the Killing vector $\chi = \partial_t + \omega \partial_\phi$ is the null generator of the horizon that is rotating with angular velocity $\omega$. Comparing it with the static case, one finds
\begin{eqnarray}\label{muprime}
    \mu^\prime  = \mu\sqrt{1 - \omega^2 l^2} \;,
\end{eqnarray}
which shows that the chemical potential transforms as the inverse of the Lorentz factor. In particular, we know that $\mu'$ corresponds to the chemical potential measured in the comoving (rotating) frame of the plasma and is related to the boundary chemical potential $\mu$ through a Lorentz factor. This relation ensures consistency with relativistic thermodynamics. Although our phase diagram is expressed in terms of $\mu$, the introduction of $\mu'$ provides a useful interpretation of density effects in rotating systems. From the gauge field in the rotating system \eqref{A0} - \eqref{AiAz}, using the relation \eqref{qmu} and assuming $\eta = 1$, one obtains the following  on-shell version of the $U(1)$ action: 
\begin{eqnarray}\label{action2}
    I_{\textsc{VF}_{\textsc{on-shell}}} =  \frac{2lLc^2 \mu^2}{g^2_5(e^{cz_h^2}-1)^2 }\gamma^4 V_{3D} \, \int_0^{\beta_s} dt  \int_\epsilon^{z_{min}}dz\, z e^{-c z^2}\left[(1+l^2\omega^2)^2 + 4l^2\omega^2f(z,\mu)\right]\;.
\end{eqnarray}

The total on-shell action in the exact soft-wall model for a rotating QCD matter is then
\begin{equation}\label{totalI}
    I_{\textsc{on-shell}} = I_{G_{\textsc{on-shell}}} + I_{\textsc{VF}_{\textsc{on-shell}}}\;, 
\end{equation}
as defined by equations \eqref{action1onshell} and \eqref{action2}. Defining the total action density by $\mathcal{E} = \frac{1}{lV_{3D}} I_{\textsc{on-shell}}$, using Eq. \eqref{totalI} one obtains
\begin{equation}\label{Es}
	\mathcal{E}_s(\varepsilon) =  \beta_s  \int_\varepsilon^{z_{min}}dz\, \frac{e^{-c z^2}}{z^5}\left[ \frac{4L^3}{\upkappa^2} + \frac{2Lc^2\mu^2}{g^2_5(e^{cz_h^2}-1)^2}\gamma^4\left((1+l^2\omega^2)^2+4l^2\omega^2f(z,\mu)\right) z^6\right]\;, 
\end{equation}
where we introduced the ultraviolet  regulator $\varepsilon$ in the integration over $z$. The regularized BH action density, without UV divergencies in the limit $\varepsilon \rightarrow 0$, is defined as the difference between the action densities of each space,  \begin{equation}\label{DeltaE}
	\bigtriangleup \mathcal{E}(\varepsilon) = \lim_{\varepsilon \rightarrow 0} \left[\mathcal{E}_{BH}(\varepsilon) - \mathcal{E}_{AdS}(\varepsilon) \right]\;. 
\end{equation}

Defining the dimensionless variables as
\begin{eqnarray}
    \bar{z}_h &=& z_h \sqrt{c}\;,\nonumber\\
    \bar{\mu} &=& \mu/\sqrt{c}\;,\nonumber\\
    \bar{q} &=& q/c^{3/2}\;,\label{varsw}
    \end{eqnarray}
from the time periods $\beta$ and $\beta_{AdS}$, see Eq. \eqref{betaAdS}, and horizon position of each space, one obtains the final expression for the regularized charged rotating BH action density: 
\begin{eqnarray}\label{deltaEsoft}
    \bigtriangleup \bar{\mathcal{E}}(\bar{\mu},\omega , \bar{z}_h) &=& \frac{e^{-\bar{z}_h^2}\pi \bar{z}_h \gamma(\omega l) } {  2 z_h^4\left(1-\frac{\bar{\mu}^2 \bar{z}_h^6}{2(e^{\bar{z}_h^2}-1)^2}\right) } \left[2(-1+\bar{z}_h^2) + e^{\bar{z}_h^2}\left(1+\frac{\bar{\mu}^2 \bar{z}_h^6}{(e^{ \bar{z}_h^2}-1)^2}\right)  + 2\bar{z}_h^4e^{\bar{z}_h^2}\text{Ei}(-\bar{z}_h^2)\right.\nonumber\\
    && \qquad\qquad\qquad\qquad\qquad-\left.\frac{\bar{\mu}^2z_h^4}{(e^{ \bar{z}_h^2}-1)^2}\bar{s}_1(\omega l, \bar{z}_h) +\frac{\bar{\mu}^4z_h^8}{(e^{ \bar{z}_h^2}-1)^4}\bar{s}_2(\omega l, \bar{z}_h)\right]\;,\nonumber\\ 
\end{eqnarray}
where $\text{Ei}(x) = - \int_{-x}^\infty e^{-t}/t\,dt $ is the exponential integral, and $\bigtriangleup \bar{\mathcal{E}} = \upkappa^2 \bigtriangleup \mathcal{E}/(L^3 c^{3/2})$ is the dimensionless action density, with the definitions
\begin{eqnarray}
    \bar{s}_1(\omega l, \bar{z}_h) &=& \gamma^4\left[ 3 + 3\omega^4l^4 - \frac{6 \omega^2 l^2}{\bar{z}_h^4}\left( 4 + 4\bar{z}_h^2 - \bar{z}_h^4\right)\right] \;,\\
    \bar{s}_2(\omega l, \bar{z}_h) &=& \frac{12  \omega^2 l^2\gamma^4}{z_h^8}\left(6 + 4\bar{z}_h^2 + \bar{z}^4_h\right)\;.
\end{eqnarray}

The regularized BH action density \eqref{DeltaE} defines the HP transitions, which correspond to deconfinement transitions via gauge/gravity duality. After computing the critical horizon positions, it defines the critical temperatures as a function of $\mu$ and $\omega l$. When $\bigtriangleup \mathcal{E}$ is positive (negative), the BH is unstable (stable), since the Gibbs free energy density ($\Phi_{\textsc{Gibbs}} = \frac{1}{\beta}\bigtriangleup \mathcal{E}$) of the AdS space is smaller (greater) than the BH one. Precisely, the phase transition occurs when
\begin{equation}\label{HPtransition}
    \bigtriangleup \bar{\mathcal{E}}(\bar{\mu},
 \omega l, \bar{z}_h) = 0  \quad  \text{at} \quad \bar{z}_h = \bar{z}_{h_c}(\bar{\mu}, \omega l)\;.
\end{equation} 
In the AdS/QCD approach, the thermal AdS space corresponds to the hadronic phase, whereas the BH phase describes the QGP. Eq. \eqref{HPtransition} does not have an analytical solution, so we must resort to numerical methods. Our goal is to perform a thorough numerical analysis of the HP transition equation \eqref{HPtransition}, identifying the relativistic rotation effects on phase transitions in the low- and high-density regimes, comparing the results with what is expected from a consistent description of the QCD phase diagram.

\section{First-order transitions for non-rotating matter}
\label{sec3}

\begin{figure}[!htb]
	\centering
    \includegraphics[scale=0.54]{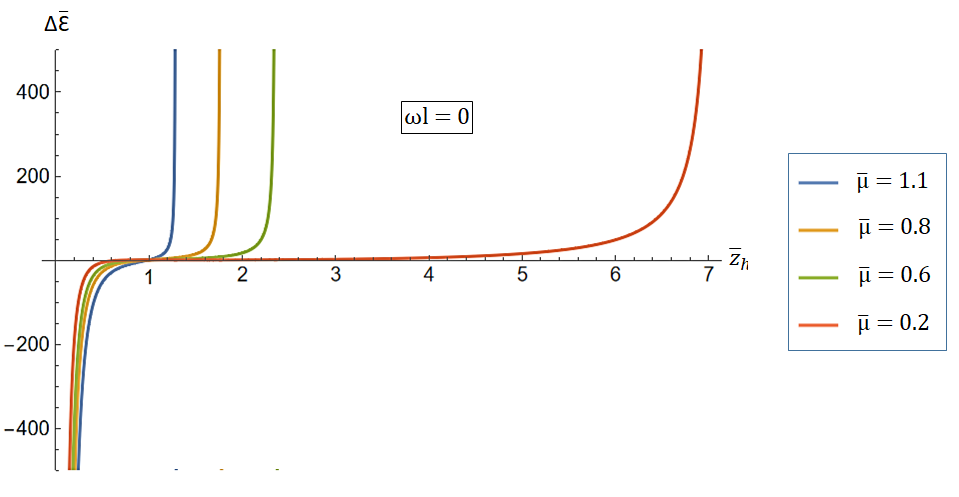}
	\caption{Action density of non-rotating charged BH as a function of the horizon position in the exact Andreev's soft wall model at different quark chemical potentials.}
    \label{fig1}
\end{figure}

\begin{figure}[!htb]
	\centering
    \includegraphics[scale=0.54]{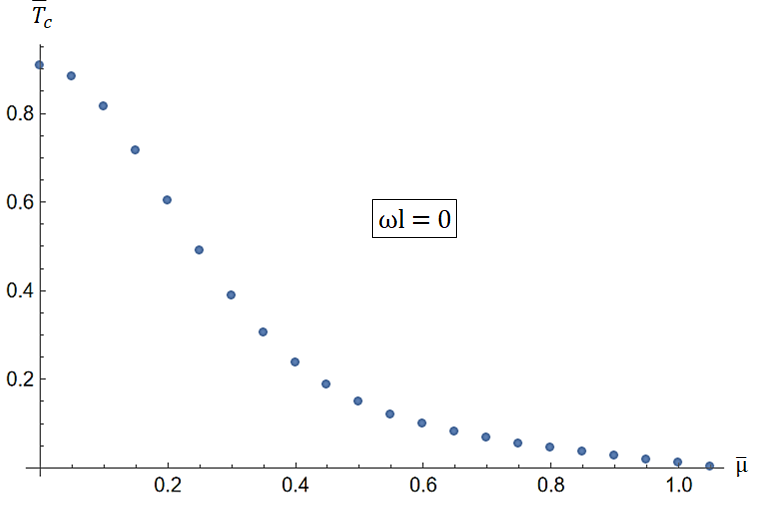}
	\caption{Phase diagram for a non-rotating QCD matter. Critical temperatures of deconfinement as a function of the quark chemical potential at $\omega l = 0$.}
    \label{fig2}
\end{figure}

The Hawking temperature is a function of $\mu$, $\omega l$, and $z_h$ together. The critical temperatures for deconfinement of rotating matter at a given density can be computed once the critical horizons are known. Using Eqs. \eqref{HTrot} and \eqref{qmu}, its dimensionless version in Andreev's exact soft wall model reads
\begin{eqnarray}\label{barT}
\bar{T}(\bar{\mu}, \omega l, \bar{z}_h) \equiv \frac{T}{\sqrt{c}}=  \frac{1}{\pi \bar{z}_h} \left(1-\frac{{\bar{\mu}}^2 z_h^6}{2(e^{\bar{z}^2_h}-1)^2}\right) \sqrt{1-\omega^2 l^2}\;. 
\end{eqnarray}
For the case without rotation, we must perform the numerical analysis of the equation
\begin{equation}\label{HPtransition2}
    \bigtriangleup \bar{\mathcal{E}}(\bar{\mu},
 0, \bar{z}_h) = 0  \quad  \text{at} \quad \bar{z}_h = \bar{z}_{h_c}(\mu)\;,
\end{equation} 
where the critical horizon appears as a function of the density.  In Fig. \ref{fig1}, we have plot $\bigtriangleup \bar{\mathcal{E}}$ as a function $z_h$ at $\omega l = 0$. As it is shown, the critical horizons ${z_h}_c$ are sensitive to the chemical potential. Table \ref{table1} lists the values of ${z_h}_c$ at different quark densities (see Appendix A), which were used to compute the corresponding critical temperatures (see Table \ref{table1}). In Fig. \ref{fig2}, we have plotted $T_c$ as a function of $\mu$ for a non-rotating matter according to Table \ref{table1}. The behavior of $T_c$ is similar to that obtained in \cite{Braga:2024nnj}. The difference here is that we do not apply the Reissner-Nordstr\"om (RN) approximation, which consists of taking the limit of small $z$ in the gauge field solution \eqref{A0final}.     

One observes that $T_c$ decreases with increasing quark density, until it reaches a maximum density beyond which no further transitions occur and the matter is always described by a plasma. Throughout the curve, the transitions are first-order transitions, in which the Gibbs free energy jumps from one phase to another as the temperature crosses $T_c$. The free energies of each phase can be calculated by holographic renormalization; see, for instance, Ref. \cite{BallonBayona:2007vp}. For a system in which all particles have zero angular momentum, there is no distinction between the type of transition in the low- and high-density regimes. In non-central heavy-ion collisions, however, the plasma formed exhibits strong vorticity, with angular velocities approaching the speed of light. Such effects must be accounted for in an accurate description of the QCD phase diagram.

\section{First-order transitions at large densities}
\label{sec4}

\begin{figure}[!htb]
	\centering
    \includegraphics[scale=0.54]{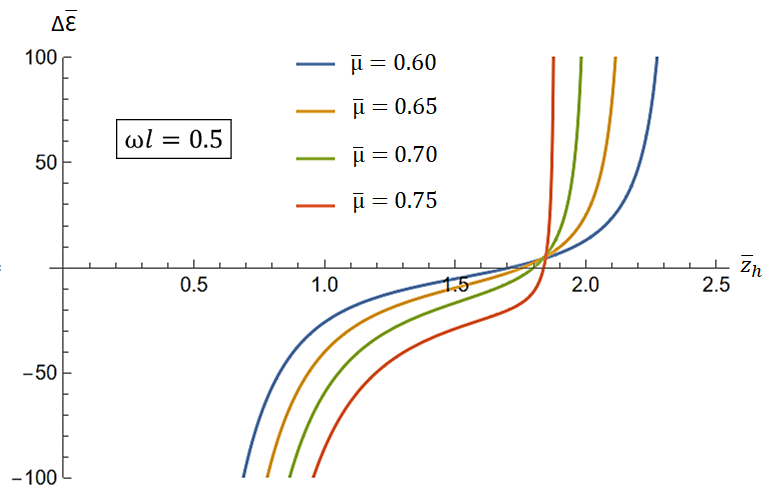}
	\caption{Action density of a charged rotating BH as a function of the horizon position in Andreev's soft wall model, at a fixed plasma rotational velocity ($\omega l = 0.5$), and different quark chemical potentials.}
    \label{fig3}
\end{figure}

At high densities, the effect of plasma rotation is observed as a decrease in the critical temperature as the plasma rotation increases \cite{Braga:2025eiz}. In addition, the maximum density value for the transition to occur also decreases with rotation. This critical density corresponds to the HP transition at zero temperature. In Ref. \cite{Braga:2025eiz}, this result was obtained using the RN approximation. The detailed behavior of this critical density as a function of the plasma rotational velocity in the exact Andreev's soft-wall model was obtained in Ref. \cite{Junqueira:2025xnx}. In this regime, the system exhibits first-order transitions at low temperatures. Between the $T_c(\mu)$ curves at different fixed angular velocities, there are narrow regions where the QGP and the hadronic matter could coexist. This phenomenon is enhanced by the fact that the maximum critical density at $T = 0$ depends on the plasma rotation. To better visualize this interpretation, see Fig. \ref{fig8} in Sec. \ref{sec6}, where we have plotted $\bar{T}_c$ curves as a function of $\bar{\mu}$ at different rotational velocities, in the low- and high-density regimes, and observe the typical first-order transitions at low temperatures. These transitions at high densities are shown in Fig. \ref{fig3}, extracted from Ref. \cite{Junqueira:2025xnx}, which are similar to the curves of Fig. \ref{fig1}, that describe first-order transitions for a non-rotating matter. For lower values of $z_h$ (higher temperatures), the QGP is always stable with a negative $\bigtriangleup \bar{\mathcal{E}}$, while for higher values of $z_h$ (smaller temperatures), the matter is always in the hadronic confined phase. This configuration differs for systems with lower densities, as we will see in Sec. \ref{sec6}.

\section{First-order transitions at zero density}
\label{sec5}
The HP transitions at zero density are computed from the numerical analysis of the equation
\begin{equation}\label{HPtransition3}
    \bigtriangleup \bar{\mathcal{E}}(0,
 \omega l, \bar{z}_h) = 0  \quad  \text{at} \quad \bar{z}_h = \bar{z}_{h_c}( \omega l)\;\;,
\end{equation} 
with the critical horizons, in principle, depending only on the plasma rotation. However, for $\bar{\mu} = 0$, the influence of the functions $\bar{s}_1(\omega l, \bar{z}_h)$ and $\bar{s}_1(\omega l, \bar{z}_h)$ is canceled out. As a consequence, the difference between the rotating system and the non-rotating one is given only by a Lorentz factor, 
\begin{equation}\label{deltaE2}
    \bigtriangleup \bar{\mathcal{E}}(0,
 \omega l, \bar{z}_h) = \gamma(\omega l) \bigtriangleup\bar{\mathcal{E}}(0,
 0, \bar{z}_h)  \;,
\end{equation}
which shows that $\bar{z}_c$ also does not depend on $\omega l$. In this case, the critical temperatures of the rotating system are given by $T_c(\omega l) = \sqrt{1 - \omega^2l^2} T_c(0)$, being $T_c(0)$ the deconfinement temperature in the static case. The inverse of the Lorentz factor is given by the Hawking temperature \eqref{HTrot} at zero density. It demonstrates that first-order transitions of Herzog's type give these transitions \cite{Herzog:2006ra}, with $T_c(\omega l)$ decreasing with rotation by a factor $1/\gamma(\omega l)$, defining the curve where the matter jumps from the hadronic phase to the deconfined one  \cite{Braga:2022yfe}. 

\begin{figure}[!htb]
	\centering
    \includegraphics[scale=0.54]{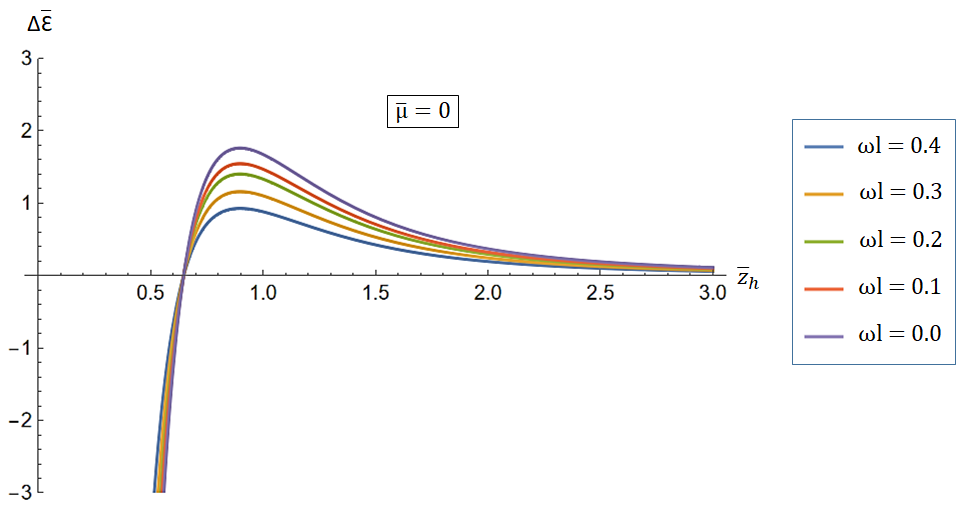}
	\caption{Action density of a rotating BH as a function of the horizon position in Andreev's soft wall model at zero density, with different rotational velocities.}
    \label{fig4}
\end{figure}

In the phase diagram, these first-order transitions occur along the temperature axis. This result is consistent with the transition predicted by EMD models at zero chemical potential \cite{Cai:2012xh}. The $\bigtriangleup\bar{\mathcal{E}}$ curves in Fig. \ref{fig4} represent this type of transition, with the BH action density crossing the $\bar{z}_h$ axis only once at the same point. In the next section, we will carefully analyze the transitions that occur at low (but not zero) densities, between first-order Herzog-type transitions at $\mu = 0$ and first-order transitions in the high-density regime that occur at relatively low temperatures, compared to those that occur at low densities, As we will see, there is a non-trivial behavior of the charged BH action densities when subjected to relativistic rotational effects, which are capable of disturbing the stability of QCD matter. Clearly, the phase transitions between the hadronic matter and the QGP in this interval will not be described by first-order transitions for relativistic rotational velocities.

\section{Phase transitions at low densities for relativistic rotational velocities}
\label{sec6}

\begin{figure}[!htb]
	\centering
    \includegraphics[scale=0.5]{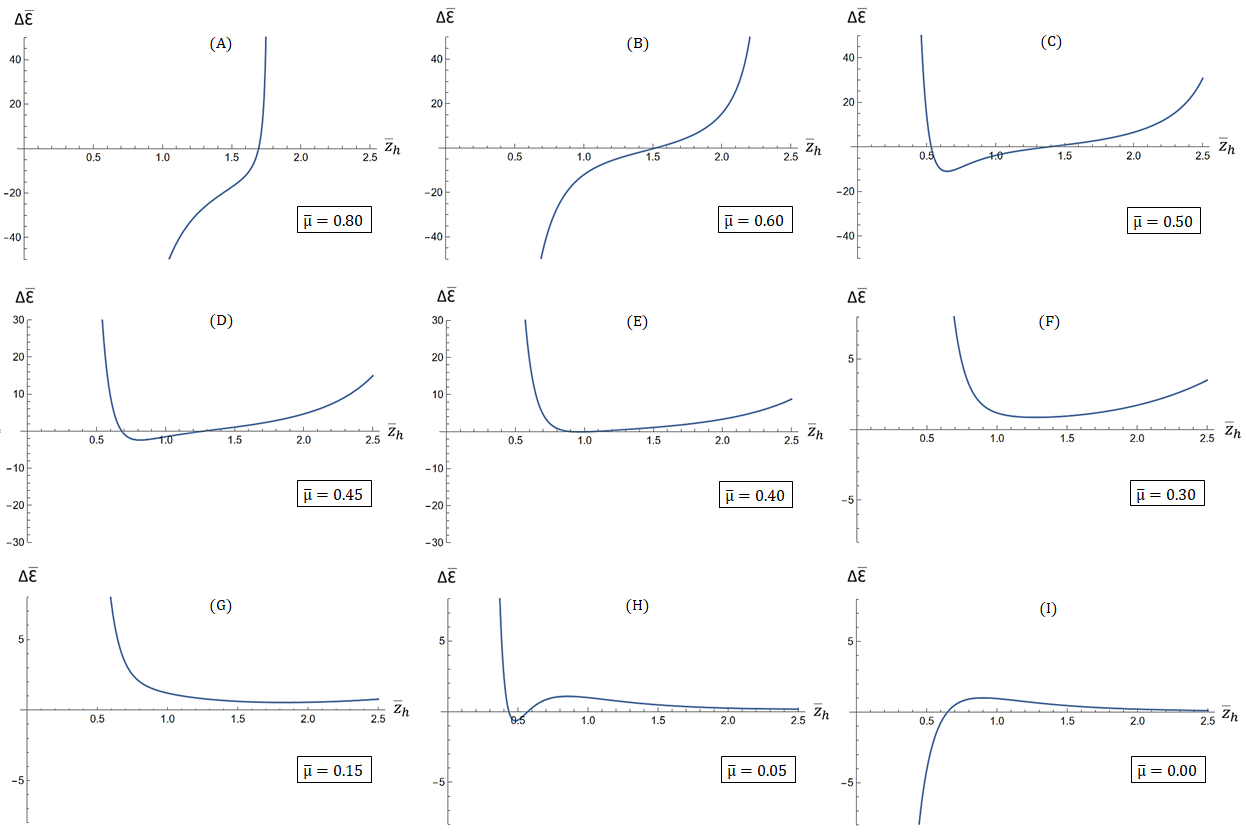}
	\caption{Phase transition at $\omega l = 0.4$. Action densities of a rotating charged BH as a function of horizon position in Andreev's soft-wall model, at different chemical potentials.}
    \label{fig5}
\end{figure}

One can observe in Fig. \ref{fig5} a distinct behavior of the BH action density curves between the first-order transition at high density, Fig. \ref{fig5}-(A), and the first-order transition of Herzog's type at zero density, Fig. \ref{fig5}-(I). As the chemical potential decreases, starting from the Fig. $\ref{sec5}$-(A), the system traverses a region in which no transitions occur, represented by the transition from Fig. \ref{sec5}-(E) to Fig. \ref{sec5}-(F). At this point, QGP reaches highly unstable limits. In Fig. \ref{sec5}-(F), the hadronic phase is always stable, since $\bigtriangleup \bar{\mathcal{E}} > 0$ independent of the temperature of the matter. We will call $\bar{\mu}_{EF}$ the critical value of the chemical potential where the transition from Fig. \ref{sec5}-(E) to (F) occurs (which we will call transition of type $E \rightarrow F$), corresponding to the exact point at which the phase transition ceases to happen, and the QCD is always in the confined phase. 

In Fig. \ref{sec5}, we plot the curves at a fixed rotational velocity $\omega l = 0.4$. The same was plotted in Fig. \ref{sec6}, but for  rotating matter with $\omega l = 0.6$. One observes that the value of $\bar{\mu}_{EF}$ depends on the rotational velocity. Table \ref{table3} contains the values of $\bar{\mu}_{EF}$ at different rotational velocities. The $E \rightarrow F$ transitions only occur for relativistic rotational velocities with  $\omega l \gtrsim 0.16$. We use Table \ref{table2} to plot $\bar{\mu}_{EF}$ as a function of $\omega l$, see Fig. \ref{fig7}. The consequence of $\bar{\mu}_{EF}$  being a function of $\omega l$ as shown in Fig. \ref{fig7}, is that there will be regions at low densities without first-order transitions, and in which plasma and hadronic matter can coexist with different angular momentum, even at high temperatures.

\begin{figure}[!htb]
	\centering
    \includegraphics[scale=0.5]{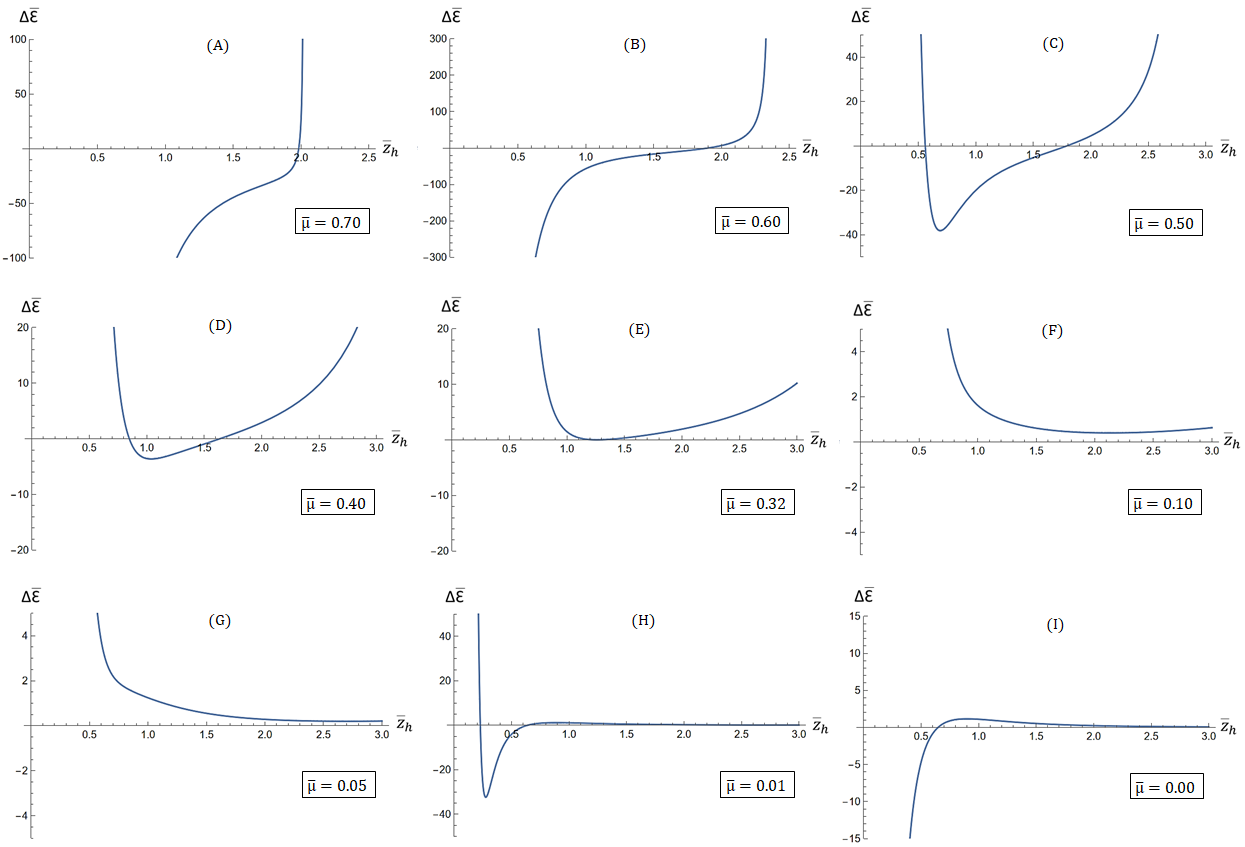}
	\caption{Phase transition at $\omega l = 0.6$. Action densities of a rotating charged BH as a function of the  horizon position, at different chemical potentials.}
    \label{fig6}
\end{figure}

\begin{figure}[!htb]
	\centering
    \includegraphics[scale=0.54]{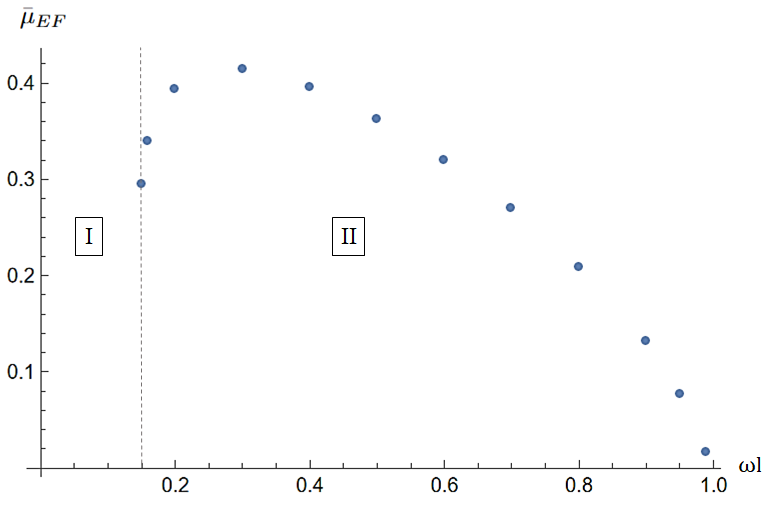}
	\caption{Minimum value of the quark chemical potential for density action transitions of type $E \rightarrow F$ as a function of the rotational velocity, with $\omega l \gtrsim 0.16$.}
    \label{fig7}
\end{figure}
\begin{figure}[!htb]
	\centering
    \includegraphics[scale=0.54]{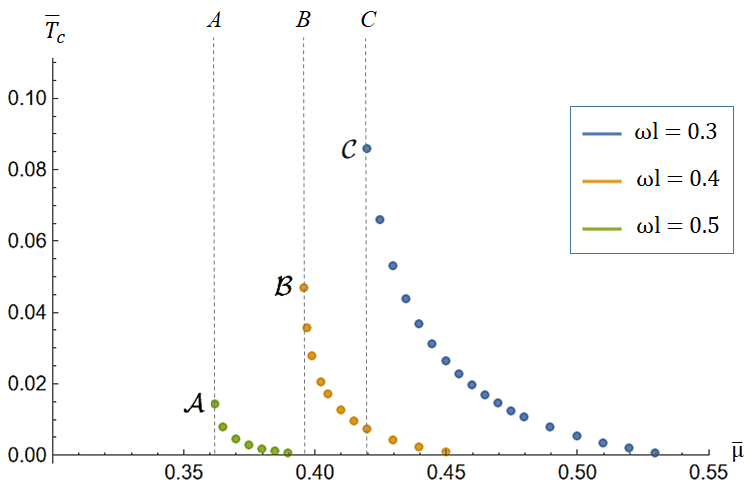}
	\caption{Critical temperatures as a function of the quark chemical potential at different angular velocities. The points $\mathcal{A}$, $\mathcal{B}$ and $\mathcal{C}$, with temperature $T_\mathcal{A}$, $T_\mathcal{B}$ and $T_\mathcal{C}$, correspond to the $E \rightarrow F$ transitions for $\bar{\mu}_\mathcal{A} = \bar{\mu}_{EF}(\omega l = 0.5)$, $\bar{\mu}_\mathcal{B} = \bar{\mu}_{EF}(\omega l = 0.4)$ and $\bar{\mu}_\mathcal{C} = \bar{\mu}_{EF}(\omega l = 0.3)$, respectively. For $\bar{\mu} \leq \bar{\mu}_{EF}$, the rotating matter with the corresponding angular momentum is always in the confined hadronic phase.}
    \label{fig8}
\end{figure}

To visualize this result, we plot the critical temperatures as a function of $\bar{\mu}$ at different rotational velocities in Fig. \ref{fig8}, according to Tables \ref{table3}, \ref{table4}, and \ref{table5}. To the left of the dashed vertical line C, the matter with rotational velocity $\omega l = 0.3$ (blue points) is always in the hadronic phase (except for a narrow region near the temperature axis where $\bar{\mu} \approx 0$, see Fig. \ref{fig5}-(H)). This dashed line corresponds to $\bar{\mu}_{EF}(\omega l = 0.3)$. Between the dashed lines B and C, there is a region at high temperatures, \textit{i. e.}, for $\bar{T} \geq \bar{T}_c(\omega l = 0.4)$, where the QGP with rotational velocity $\omega l = 0.4$ (yellow points) coexists with the hadronic matter with $\omega l = 0.3$. The vertical dashed line B corresponds to $\bar{\mu}_{EF}(\omega l = 0.4)$, see Table 3. The same argument applies to the regions between the vertical dashed lines A, B, and C.  This phase-mixing at low energies occurs even for small values of $\bar{\mu}$, as shown in Fig. \ref{fig7}. As $\bar{\mu}$ tends to zero, the mixture will occur due to states that rotate at velocities comparable to the speed of light.

Also from Fig. \ref{fig7}, one concludes that the coexistence between the two phases at high temperatures is limited to a defined region at low densities, until it reaches the maximum value of $\bar{\mu}_{EF}$, \textit{i. e.}, for $\bar{\mu} \approx \bar{\mu}_{EF}(\omega l = 0.3) = 0.414 $. The most critical density is given by the value of $\bar{\mu}$ at $T=0$ for non-rotating matter, which is $\bar{\mu}(T\approx 0) \approx 1.067$. For $\bar{\mu} \geq 1.067$, there are no phase transitions. For $0.414 \leq\bar{\mu} \leq 1.067$, the system exhibits first-order transitions. In principle, these transitions also allow the coexistence of the two phases between the $T_c$ curves at different rotational velocities, even at very low temperatures, since the values of the maximum critical density at $T = 0$ decrease
with $\omega l$. This type of coexistence can occur only at temperatures below the $T_c(\omega l = 0)$ curve, and it is not expected to appear prominently in the QCD phase diagram. At lower temperatures, there is little energy available for particle kinetics, as particle interactions in highly dense states consume a large fraction of the energy. In this case, the transitions will be dominated by first-order transitions for temperatures slightly lower than $T_c(\omega l = 0)$, and the coexistence between the phases at high densities should occur in an extremely narrow region at very low temperatures, see Fig. \ref{fig8} -- between the blue and yellow $T_c$ curves (where the phases are mixed), $T < 0.08$, approximately. It decreases even further as rotational velocities increase.  We will discuss the non-relativistic limit in the next section. 

On the other hand, for $\bar{\mu} \leq \text{max}(\bar{\mu}_{EF}) \approx 0.414$, the effect of relativistic rotations is expected to be quite significant in phase transitions. At high temperatures and low densities, a large amount of energy is available for particle kinetics. In this case, the coexistence of the QGP and hadronic matter will be significant, and the phase transitions will not be first-order. In Fig. \ref{fig8}, we have divided the space by the dashed lines A, B, and C. In practice, we could perform infinite subdivisions in the region $\bar{\mu} \leq \text{max}(\bar{\mu}_{EF})$. For each state at a fixed quark density in this region, the total Gibbs free energy must be given by an infinite sum over all plasma and hadron states, coexisting with different angular momentum. Therefore, the transitions must occur smoothly pointwise, not through a jump from one phase to another. As the temperature increases, we should observe a smooth (crossover) transition, governed by the behavior of the negative QCD coupling constant. Strong interactions tend to weaken at the limit of extremely high temperatures, indicating a strong instability of hadronic matter. This interpretation is consistent with the instability predicted by the specific heat analysis, as we will see in the next section. For this reason, we disregard the second phase transition that appears in some intermediate density regimes --- see, for instance, Fig. \ref{fig6}-(C) and (D) --- which indicate the stability of hadronic matter at extreme temperatures, at energy regimes where the free energies are not determinant for describing the stability of QCD matter. Complementing the analysis, as the temperature decreases, the $E \rightarrow F$ transitions should not take place, given by states with much lower rotational velocities (when compared to the speed of light), in which the phase transitions should be described by the critical temperature $T_c(\omega l \approx 0)$, at which hadronic matter is always stable at lower temperatures. In short, the smooth transitions in the QCD diagram should appear in an intermediate low-density region between the low and extremely high-temperature regimes, understood as a result of relativistic rotations.

\subsection{Gibbs free energy, entropy and specific heat from holographic renormalization}
\label{sec61}
In the previous section, we have qualitatively discussed smooth transitions in the low-density regime. A single state at a given angular velocity undergoes an HP (first order) transition defined by its critical temperature curve $T_c(\omega l)$, according to the density of the medium. It jumps from the hadronic phase to the plasma phase as the temperature crosses $T_c$. The transition type is determined by analyzing thermodynamic properties derived from the Gibbs free energy, such as entropy and specific heat. As an illustrative example, we will consider point $\mathcal{B} \equiv  (\bar{T}_\mathcal{B}, \bar{\mu}_\mathcal{B})$, see Fig. \ref{fig8}. For a state with angular momentum $\omega l = 0.4$ at $\bar{\mu}_\mathcal{B} = 0.396$, the HP transition will occur at $\bar{T}_{\mathcal{B}} =0.0466958$. In the presence of rotation, the system possesses an extra degree of freedom: the angular momentum. In the analysis of the $2D$ phase diagram, described by $T_c(\mu)$, the angular momentum axis is implicit. The complete phase diagram is three-dimensional, described by the surface $T_c(\mu, \omega l)$. For a system with $N_p$ particles, assuming an ensemble with large $N_p$, each point $(T,\mu)$ can be occupied by multiple states with different angular momentum. The combination of states in different phases at a given chemical potential can be determined from the analysis of the $E \rightarrow F$ transitions.   

By analyzing the point $\mathcal{B}$, for example, as the matter is always in the hadronic phase for $\bar{\mu} \leq \bar{\mu}_{EF}$ at a given rotational velocity, one concludes that this point is composed of a combination of hadronic states with $0.4 \leq \omega l \leq 0.3 $ (assuming the minimum of the rotational velocity given by $\omega l = 0.3$ as in Fig. \ref{fig8}) with QGP states with $0.4 \leq \omega l < 1$, since $\bar{\mu}_{EF}$ decreases with $\omega l$ in this region, see Fig. \ref{fig7}. The same combination is valid  along the dashed line B, see Fig. \ref{fig8}, indicating the coexistence of hadronic and plasma states at high temperatures. To obtain the Gibbs free energy of a single state, we must apply holographic renormalization to eliminate UV divergencies of each phase. The holographic renormalization consists of introducing a boundary Gibbons-Hawking (GH) term into the action, with its corresponding surface counterterms. In the soft-wall AdS approach, see \cite{Arutyunov:1998ve, BallonBayona:2007vp}, the GH action is given by
\begin{equation}
I_{GH} = -  \frac{1}{\kappa^2} \int_{\partial M} d^4x \sqrt{h}\,  u(\Phi) K\;, 
\end{equation}
where $u(\Phi)$ is a function of the dilaton background field,  $K$ is the trace of the extrinsic curvature of the boundary, and $h$ is the determinant of the boundary induced metric $h_{\mu\nu}$, such that 
\begin{equation}
K = \nabla n^a = \frac{1}{\sqrt{g}} \partial_a \left( \sqrt{g} \,n^a \right)\;,
\end{equation}
where $n^a$ is a unitary vector normal to the boundary. For the BH AdS geometry,  
\begin{equation}
n^a = (-z/R,0,0,0,0) \quad \text{with} \quad  \sqrt{h} = \frac{R^4}{z^4} \;,
\end{equation}
and, for the thermal AdS one,
\begin{equation}
n^a = (-z f(z,q)/R,0,0,0,0)\quad \text{with} \quad  \sqrt{h} = \frac{\sqrt{f(z,q)} R^4}{z^4} \;.
\end{equation}

In this case, the surface counterterm action takes the form
\begin{equation}
I_{ct} = \frac{1}{\kappa^2} \int_{\partial M} \sqrt{h} \,u(\Phi) F(R, \tilde{R}, \nabla \tilde{R})\;,
\end{equation}
where $F$ is a finite series of diffeomorphism invariants constructed from the AdS radius $R$
and the boundary induced curvature $\tilde{R}$. The functions $u$ and $F$ are fixed by the terms that eliminate the UV divergencies of the total renormalized action, 
\begin{equation}\label{Ihr}
I_{total}^{(hr)} = I_{\text{bulk}} + I_{\text{surface}} \quad \text{with} \quad I_{\text{surface}} = I_{GH}+I_{ct}\;,
\end{equation}
wherein $I_{bulk} = I_{\textsc{on-shell}}$, see Eq. \eqref{totalI}, which yields
\begin{equation}\label{uF}
u = -1 + 2\Phi + \Phi^2 \text{log}\,\Phi \quad \text{and} \quad F = \frac{3}{R}\;.
\end{equation}

From the result \eqref{uF}, one obtains the following action densities 
\begin{eqnarray}
\mathcal{E}_{surface}^{AdS} &=& \frac{L^3 \beta_{AdS}}{\kappa^2} \left[ -\frac{1}{\epsilon^4} + \frac{2c}{\epsilon^2} + c^2\text{log}(c\epsilon^2)\right] \;,\label{IsurfAdS}\\
\mathcal{E}_{surface}^{BH} &=& \frac{L^3 \beta}{\kappa^2} \left[ -\frac{1}{\epsilon^4} + \frac{2c}{\epsilon^2} + c^2\text{log}(c\epsilon^2)+ \frac{1}{2 z_h}\right]\;.\label{IsurfBH}
\end{eqnarray}

By replacing \eqref{IsurfAdS} and \eqref{IsurfBH} into Eq. \eqref{Ihr}, and using \eqref{Es} for the bulk actions in each space, one obtains the final renormalized action densities in their dimensionless versions:
\begin{eqnarray}\label{EhrAdS}
\!\!\!\!\!\!\!\!\!\!\bar{\mathcal{E}}_{AdS}^{(hr)} &=& \frac{L^3 \beta}{2\kappa^2 z_h^4} \left[(3-2\gamma_E)z_h^4 + 3q^2\gamma^4 z_h^4 + 3q^2l^2 \omega^2 \gamma^4(-8+6z_h^4+l^2\omega^2 z_h^4)- 24l^2\omega^2q^4\gamma^4(-3+z_h^2)\right]\;,
\end{eqnarray}   
and 
\begin{eqnarray}\label{EhrBH}
\!\!\!\!\!\!\!\!\!\!\!\!\bar{\mathcal{E}}_{BH}^{(hr)} &=& \frac{L^3 \beta}{2\kappa^2 z_h^4} e^{-z_h^2}\left[ 2(-1+z_h^2)-3q^2z_h^4\gamma^4-3q^2l^4\omega^4z_h^4\gamma^4-6q^2l^2\omega^2\gamma^4(-4-4z_h^2+z_h^4)-\right.\nonumber\\
&&-\left. 12l^2\omega^2q^4z_h^4\gamma^4(6+4z_h^2+z_h^4)+2e^{z_h^2}+z_h^4 e^{z_h^2}(3-2\gamma_E+q^2z_h^2)+3q^2\gamma^4 z_h^4e^{z_h^2}\right.\nonumber\\
&&+\left. 3q^2l^2\omega^2\gamma^4e^{z_h^2}(-8+6z_h^4+l^2\omega^2z_h^4)+24q^4l^2\omega^2\gamma^4 e^{z_h^2}(-3 + z_h^2) + 2z_h^4 e^{z_h^2} \text{Ei}(-z_h^2)\right]\;,
\,
\end{eqnarray}
where $\gamma_E = 0.577216...$ is the Euler gamma constant.

The dimensionless Gibbs free energy ($G$), entropy ($S$), and specific heat ($C_V$) of a system can be computed using the expressions\footnote{It is implicit that these thermodynamic quantities are densities, divided by the radius $l$ and the spatial factor volume $V_{3D}$.}
\begin{eqnarray} \label{GSCv}
   G = \frac{1}{\beta} \bar{\mathcal{E}}\;, \quad  S = - \left(\frac{\partial G}{\partial \bar{T}}\right)_{\bar{Q}}\;, \quad \text{and} \quad C_V = \bar{T}\left(\frac{\partial S}{\partial \bar{T}}\right)_{\bar{Q}}\;.
\end{eqnarray}
Since $\bar{\mathcal{E}}^{(hr)}_{AdS}$ has been renormalized and is free of UV divergences, we can use $\beta_{AdS} = \beta$ in the expression for $G$ above. This way, for a single state with a given angular velocity, the Gibbs free energy for the confined and plasma phases is given by, respectively,
\begin{eqnarray}\label{Gs}
    G_{AdS}(\bar{T}, \omega l) = \frac{1}{\bar{\beta}} \bar{\mathcal{E}}^{(hr)}_{AdS}(\bar{T}, \omega l)\;,\quad \text{and} \quad  G_{BH}(\bar{T}, \omega l) = \frac{1}{\bar{\beta}} \bar{\mathcal{E}}^{(hr)}_{BH}(\bar{T}, \omega l)\; 
\end{eqnarray}

\begin{figure}[!htb]
	\centering
    \includegraphics[scale=0.5]{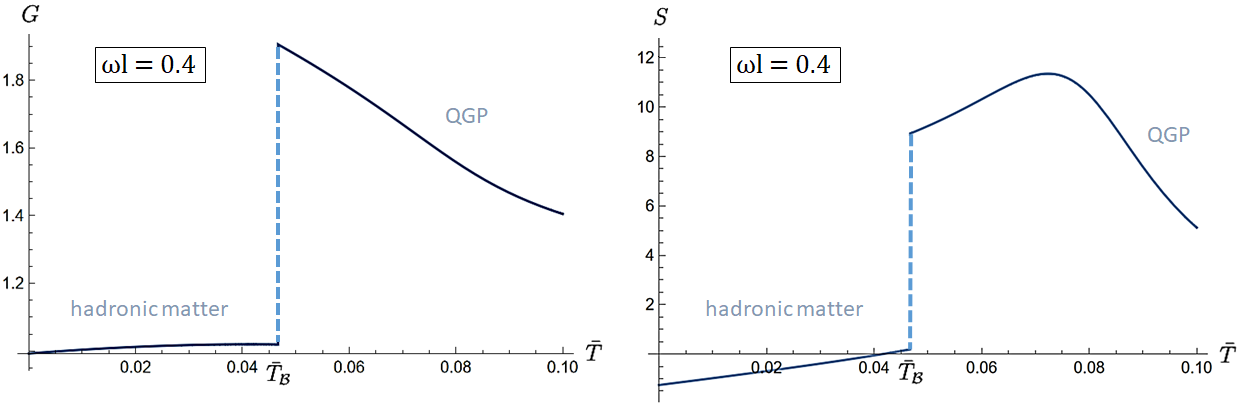}
	\caption{\textit{First order transition.} Gibbs free energy ($G$) and entropy ($S$) as a function of $\bar{T}$ at $\bar{\mu} = 0.396$, for a single state with rotational angular velocity $\omega l = 0.4 $. The temperature $T_\mathcal{B}$ represents the critical temperature $\bar{T}_c (\omega l = 0.4)=0.0466958$, see Table \ref{table4}.}
    \label{fig9}
\end{figure}

\begin{figure}[!htb]
	\centering
    \includegraphics[scale=0.54]{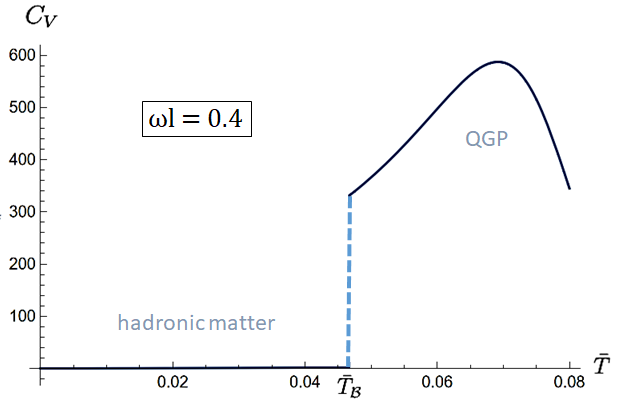}
	\caption{Specific heat ($C_V$) as a function of $\bar{T}$ for a single state with rotational velocity $\omega l = 0 .4$ at $\bar{\mu} = 0.396$.}
    \label{fig10}
\end{figure}

To obtain the Gibbs energies as a function of $\bar{T}$ and $\omega l$, we must find the horizon function $\bar{z}_h(T, \omega l)$, which can be obtained from the Hawking temperature \eqref{HTrot}. The sixth-order polynomial in $\bar{z}_h$, which comes from this equation, does not have an analytical solution. To overcome this problem, we can expand $T(\bar{z}_h, \omega l)$, at a fixed rotational velocity, around a given $\bar{z}_{h0}$, in powers of $(\bar{z}_h - \bar{z}_{h 0})^n$ up to a given order that has an analytical solution. Starting from point $\mathcal{B}$, we must take $\bar{z}_{h0} = \bar{z}_{hc}(\omega l = 0.4) = 0.986429 $. Choosing $n=3$, one can obtain an expression for $\bar{z}_h(\bar{T}, \omega l)$ and then replace it into equations \eqref{EhrAdS} and \eqref{EhrBH}. Having done this, one can apply Eq. \eqref{Gs} to obtain the curves of $\bar{G}_{AdS}(\bar{T}, 0.4)$ and $\bar{G}_{BH}(\bar{T}, 0.4)$, for $\bar{T} \leq T_\mathcal{B}$ and $\bar{T} > T_\mathcal{B}$, respectively, in a region where $\vert \bar{z}_h - \bar{z}_{h0}\vert < 1$. The curves of $S(\bar{T}, 0.4)$ and $C_V(\bar{T}, 0.4)$ are automatically obtained by applying Eq. \eqref{GSCv}. These curves are plotted in Figs. \ref{fig9} and \ref{fig10}, which represent a first order transition at $\bar{\mu}_\mathcal{B}$, as the temperature of the state with rotational velocity $\omega l =0.4$ passes through the critical temperature $\bar{T}_\mathcal{B} = T_c(\omega l = 0.4)$. This type of phase transition is general to HP transitions in a system composed of a single state with a fixed angular momentum.

Now, if we consider an ensemble with $N_p$ particles, the total free energy Gibbs for any point along the dashed line B at $\bar{\mu} = \bar{\mu}_\mathcal{B}$, see Fig. \ref{fig8}, will be given by a combination of hadronic states with $0.3 \leq \omega l \leq 0.4$ (we are assuming the minimum for the rotational velocity as $\omega l = 0.3$, according to the values analyzed in Fig \ref{fig8}) and QGP states with $0.4\leq \omega < 1$. In this region, $\bar{\mu}_{EF}$ always decreases with $\omega l$. $ E \rightarrow F$ transitions determine such a combination of states in different phases. As an illustrative example, we will consider a system composed of an ensemble of particles that can assume rotational velocities in the range $0.3 \leq \omega l \leq 0.8$. This way, assuming that discrete levels describe the angular momentum, one can write 
\begin{equation}
  (\omega l)_{\text{max}} - (\omega l)_{\text{min}} = N_p a_\omega\;,  \end{equation}
being $a_\omega$ the average interval between two neighboring levels of angular momentum, multiplied by $l$. In the large $N_p$ limit, one has $a_\omega \approx d(\omega l)$, such that the total free Gibbs energy for this type of ensemble takes the form  
\begin{eqnarray}\label{GNp}
    G_{N_p}(\bar{T}) =\frac{1}{a_\omega} \left[\int_{0.3}^{0.4}  G_{AdS}(\bar{T}, \omega l) d(\omega l) + \int_{0.4}^{0.8}  G_{BH}(\bar{T}, \omega l) d(\omega l) \right]\;.
\end{eqnarray}

\begin{figure}[!htb]
	\centering
    \includegraphics[scale=0.5]{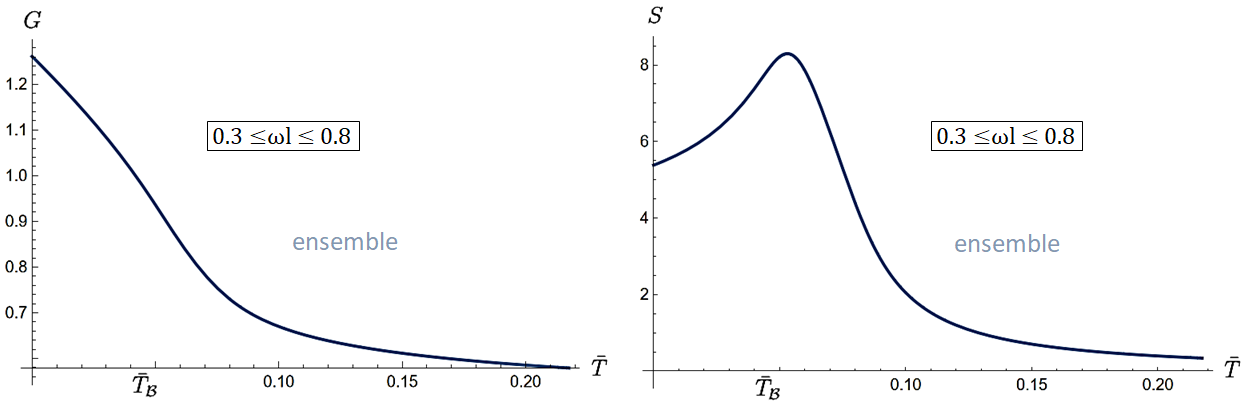}
	\caption{\textit{Smooth transition}. Gibbs free energy ($G_{N_p}^{(a)}$) and entropy ($S_{N_p}^{(a)}$) as a function of $\bar{T}$ at $\bar{\mu} = 0.396$, for an ensemble of states with distinct angular momentum, for rotational angular velocities varying from $\omega l = 0.3$ to $\omega l = 0.4 $. Again, $T_\mathcal{B} =\bar{T}_c (\omega l = 0.4)=0.0466958$. These curves represent transitions that can occur along the dashed line $B$ defined in Fig. \ref{fig8}.  }
    \label{fig11}
\end{figure}

\begin{figure}[!htb]
	\centering
    \includegraphics[scale=0.6]{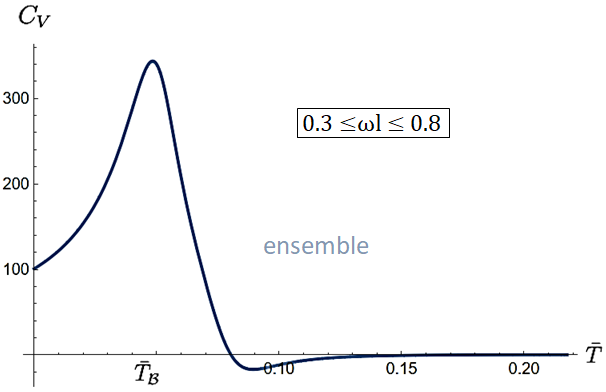}
	\caption{Specific heat ($C_{V N_p}^{(a)}$) as a function of $\bar{T}$ for the same ensemble of states with rotational velocity varying from $\omega l = 0 .3$ to $\omega l = 0 .8$,  at $\bar{\mu} = 0.396$.}
    \label{fig12}
\end{figure}

Starting again from point $\mathcal{B}$, one can plot $G_{N_p}(T)$ for temperatures varying along the dashed line B ($\bar{\mu} =\bar{\mu}_\mathcal{B}$), around the temperature $T_\mathcal{B}$. Using Eq. \eqref{GNp}, we can apply Eq. \eqref{GSCv} to plot the curves for the entropy and specific heat of the ensemble as functions of $\bar{T}$, see Figs. \ref{fig11} and \ref{fig12}. To plot these curves, we have defined 
\begin{eqnarray}\label{G(a)}
  G_{Np}^{(a)} = a_\omega G_{Np}\;, \quad  S_{Np}^{(a)} = a_\omega S_{Np}\;, \quad \text{and} \quad C_{V N_p}^{(a)} = a_\omega C_{V N_p}\;,  
\end{eqnarray}
to subtract the numerical influence of the parameter $a_\omega$, which does not affect the smooth behavior of the curves. As predicted by the qualitative analysis carried out in the previous section, the first-order transition that occurs for a single state at a given angular momentum is replaced by a smooth transition for the ensemble, without any discontinuity as the temperature varies along the dashed line B of Fig. \ref{fig8}. 

\begin{figure}[!htb]
	\centering
    \includegraphics[scale=0.54]{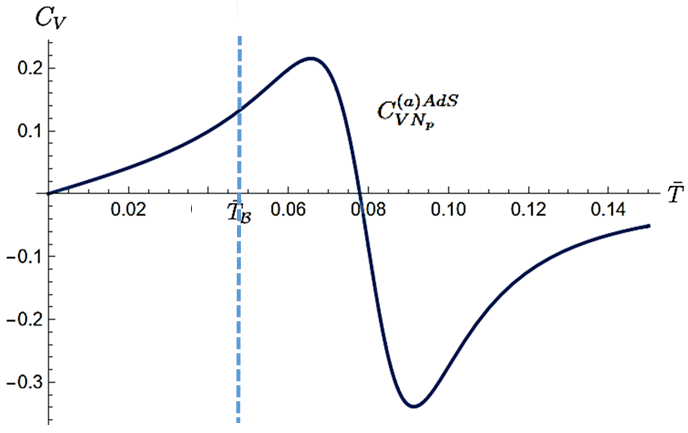}
	\caption{Specific heat ($C_{V N_p}^{(a) AdS}$) as a function of $\bar{T}$ for the ensemble composed of states in the hadronic phase with rotational velocity varying from $\omega l = 0 .3$ to $\omega l = 0 .8$,  at $\bar{\mu} = 0.396$.}
    \label{fig13}
\end{figure}

\begin{figure}[!htb]
	\centering
    \includegraphics[scale=0.54]{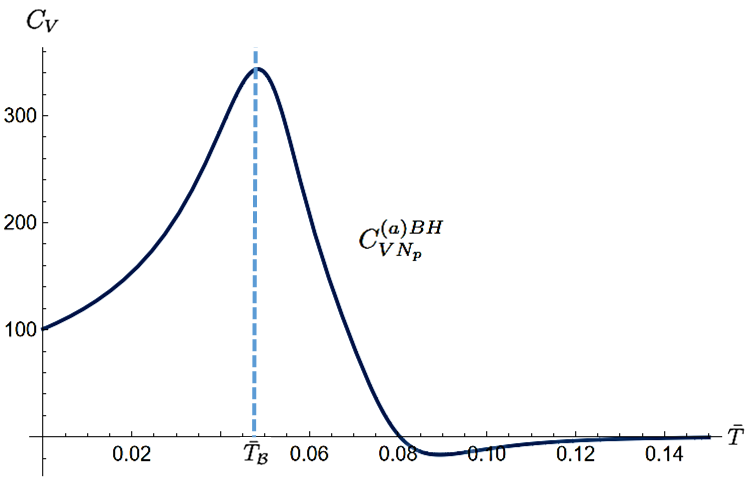}
	\caption{Specific heat ($C_{V N_p}^{(a) BH}$) as a function of $\bar{T}$ for the ensemble composed of states in the plasma phase with rotational velocity varying from $\omega l = 0 .3$ to $\omega l = 0 .8$,  at $\bar{\mu} = 0.396$.}
    \label{fig14}
\end{figure}

This result was expected, since the same combination of hadronic and plasma states occurs at $\bar{\mu} = \bar{\mu}_\mathbf{B}$, along the entire line B (including extreme high temperatures), defined by the same range of angular momentum for each phase, according to $E \rightarrow F$ transitions. This type of smooth transition should be observed only in the low-energy regime, up to the critical point $(\bar{T}_{CP}, \bar{\mu}_{CP})$, from which this type of coexistence between phases ceases to occur. For $\bar{\mu} > \bar{\mu}_{CP}$, the system is characterized by a first-order transition through $T_c(\omega l \approx 0)$, where the matter jumps from the hadronic phase to the deconfined one. In the low-density regime, the mixed phase at lower temperatures is expected to undergo a smooth transition to a pure QGP phase as the hadronic matter becomes increasingly unstable at extremely high temperatures. Assuming that the holographic AdS/QCD model consistently represents a gravitational dual of strongly interacting QCD matter, as the temperature increases continuously, the gauge coupling tends to decrease until the point where the system will be described by free particles, with $g^2 \rightarrow 0$, being dominated by the plasma phase. This interpretation is consistent with the hadronic instability indicated by the specific heat of the ensemble of states in the confined phase. By defining
\begin{eqnarray}
    G_{N_p}^{AdS}(\bar{T}) &=&\frac{1}{a_\omega} \int_{0.3}^{0.4}  G_{AdS}(\bar{T}, \omega l) d(\omega l) \;,\label{GNpAdS}\\
     G_{N_p}^{BH}(\bar{T}) &=&\frac{1}{a_\omega} \int_{0.4}^{0.8}  G_{BH}(\bar{T}, \omega l) d(\omega l) \;,\label{GNpBH}
\end{eqnarray}
one can use equations \eqref{GNpAdS} and \eqref{GNpBH} to obtain the curves for $  C_{V N_p}^{(a) AdS}$ and $  C_{V N_p}^{(a) BH}$ as functions of $\bar{T}$, with these quantities defined by equations \eqref{GSCv} and \eqref{G(a)}, see Figs. \ref{fig13} and \ref{fig14}. Particularly, Fig. \ref{fig13} shows that the specific heat of hadronic states becomes negative as the temperature increases, starting from the point with temperature $T_\mathcal{B}$, where the method used is reliable. This negative specific heat indicates a strong process of energy loss involving the particles of the system. In the gauge side of the duality, it represents the loss of mass of hadronic matter, which can occur through the emission of photons directly from hadrons in heavy-ion collisions, see \cite{Shuryak:1978ij, Peitzmann:2001mz}. This behavior complies with the negative QCD $\beta$-function, which establishes the weakening of the strong interactions as the temperature increases. This way, the matter will undergo a smooth transition in the low-density regime, from a mixed phase (formed by hadronic and plasma states with distinct angular momentum) to a phase dominated by QGP states, as the hadrons become increasingly unstable at higher temperatures.

\begin{figure}[!htb]
	\centering
    \includegraphics[scale=0.54]{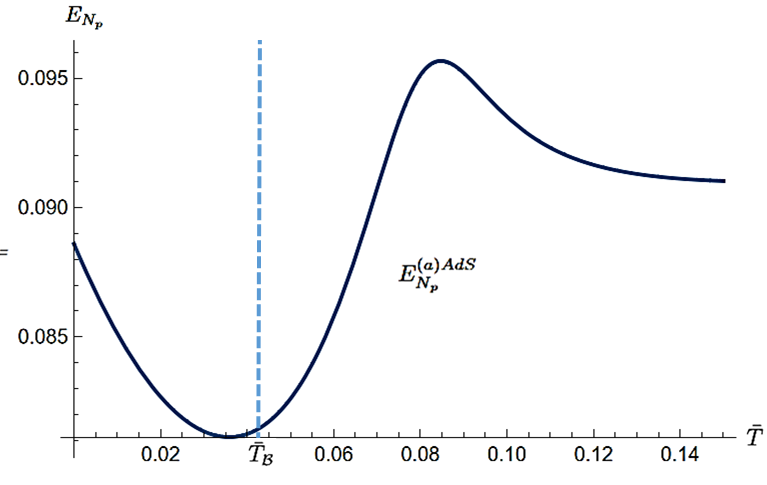}
	\caption{Total energy for plasma states in the grand canonical ensemble ($E_{N_p}^{(a) AdS}$), for rotational velocities varying from $\omega l =0.3$ to $\omega l = 0.8$, around the point $B$ of Fig. \ref{fig8}.}
    \label{fig15}
\end{figure}

\begin{figure}[!htb]
	\centering
    \includegraphics[scale=0.54]{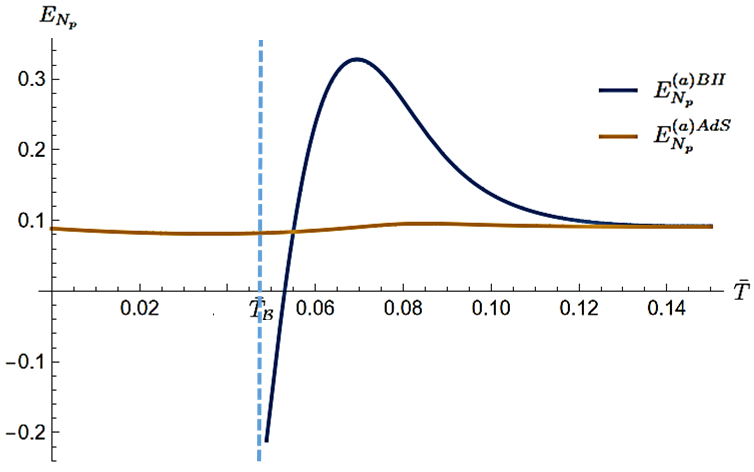}
	\caption{Total energy for plasma states ($E_{N_p}^{(a) AdS}$) compared with the total black hole energy ($E_{N_p}^{(a) BH}$) that describes QGP states, in the grand canonical ensemble, for rotational velocities varying from $\omega l =0.3$ to $\omega l = 0.8$, around the point $B$ of Fig. \ref{fig8}.}
    \label{fig16}
\end{figure}

In order to visualize this dominance of QGP states at higher temperatures, we will consider the system being described by a grand-canonical ensemble, with the total energy of the system given by 
\begin{eqnarray}
E= G + \omega J + \bar{\mu} Q_e+ \bar{T} S\;,   
\end{eqnarray}
where 
\begin{eqnarray}
    J = - \left(\frac{\partial G}{\partial \omega}\right)_{\bar{T},\bar{\mu}} \quad \text{and} \quad  Q_e = - \left(\frac{\partial G}{\partial \bar{\mu}}\right)_{\bar{T},\omega}
\end{eqnarray}
are the averages of the total angular momentum and total charge, respectively. For an expansion of $\bar{T}$ around the point $\mathcal{B} = (\bar{T}_\mathcal{B}, \bar{\mu}_{\mathcal{B}})$, along the dashed line B, see Fig. \ref{fig8}, one can use equations \eqref{GNpAdS} and \eqref{GNpBH}, together with Eq. \eqref{G(a)}, to obtain the curves of $E_{N_p}^{(a)} = a_\omega E_{N_p}$ for the states in the plasma and hadronic phases. The results are shown in Figs. \ref{fig15} and \ref{fig16}. From Fig. \ref{fig15}, one observes that as the temperature increases, the total energy of the hadrons reaches a maximum value, until it begins to decrease. This decrease corresponds to the instability of hadronic matter, which is consistently described by its negative specific heat in this region, indicating a loss of mass of these states at higher temperatures.

By comparing the total hadronic energy in the grand-canonical ensemble with the QGP one, see Fig. \ref{fig16}, one observes the energetic dominance of plasma states over hadronic ones as the temperatures increase (starting from $\bar{T} = T_\mathcal{B}$). It is interesting to note that the QGP total energy  also begins to decrease in the high-temperature regime, which is explained by its negative specific heat at this temperature range, see Fig. \ref{fig14}. This energy loss can be interpreted as the emission of photons via the Hawking radiation effect from the rotating black hole. Phenomenologically, this holographic description is consistent with the photon production by the QGP in heavy-ion collisions \cite{Wang:2001xh, PHENIX:2008uif}. In any case, this energy loss is not sufficient to destroy the dominance of plasma states over hadronic matter at high temperatures, as demonstrated by the analysis of Fig. \ref{fig16}. Since we are applying an approximation method, the analysis of infinitesimal variations in thermodynamic quantities cannot be done analytically. However, these total energies must obey the first law 
\begin{equation}
dE = \mu dQ_e + T dS + \omega dJ,
\end{equation}
which can be demonstrated from the application of a Cardy-like formula for rotating black holes,  see  \cite{BravoGaete:2017dso}. This result is generally valid in the context of gauge/gravity duality for black holes with a planar horizon. 

\subsection{First-order transitions in the non-relativistic limit}
\label{sec6.2}

\begin{figure}[!htb]
	\centering
    \includegraphics[scale=0.5]{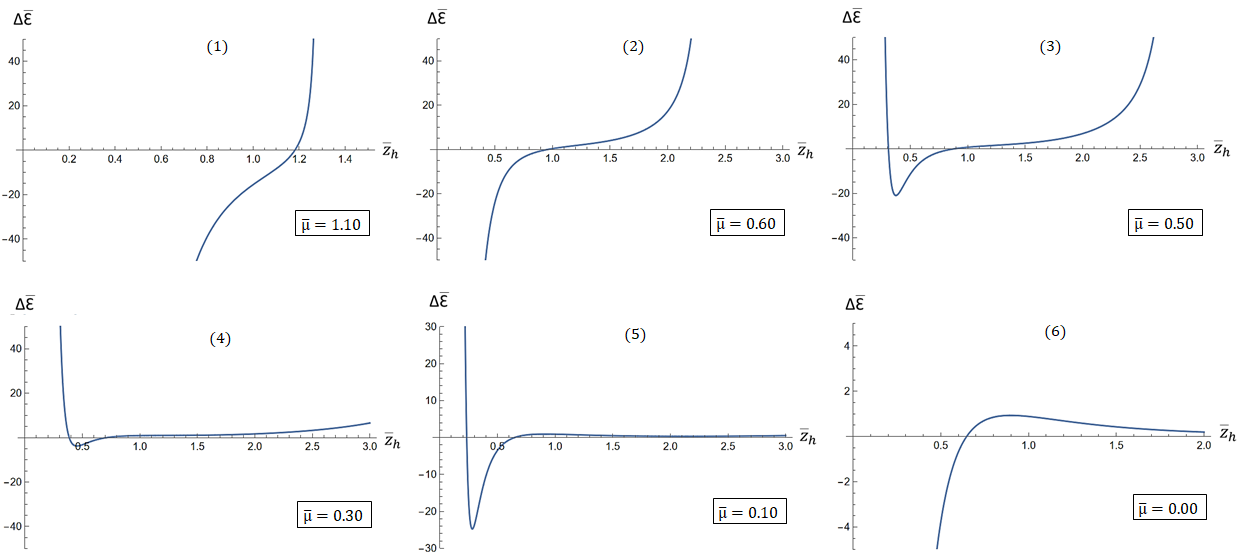}
	\caption{Phase transition at $\omega l = 0.1$. Action densities of a rotating charged BH as a function of the horizon position, at different chemical potentials.}
    \label{fig17}
\end{figure}

\begin{figure}[!htb]
	\centering
    \includegraphics[scale=0.5]{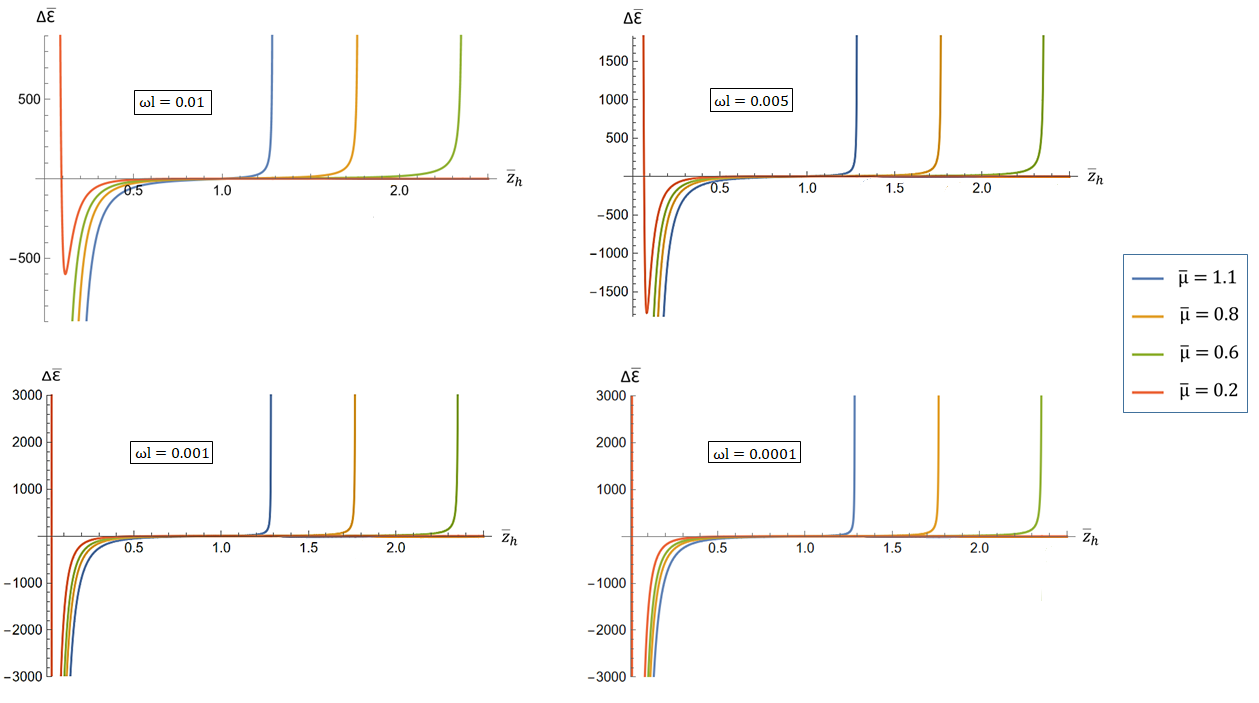}
	\caption{Non-relativistic limit of charged BH action densities, at different chemical potentials.}
    \label{fig18}
\end{figure}

We have split Fig. \ref{fig7} into regions I and II. The dashed vertical line in Fig. \ref{fig7} corresponds to $\omega l = 0.16$. Region II is characterized by $E \rightarrow F$ transitions, see Figs. \ref{fig5} and \ref{fig6}, which only occur for rotational velocities greater than $16\%$ the speed of light. For $\omega l = 0.16$, this type of transitions do not happen, that is, there is no critical density $\bar{\mu}_{EF}$ at which first-order transitions cease to exist, from which matter is always described by the confined phase for $\bar{\mu} \leq \bar{\mu}_{EF}$. Instead, the action densities are described by curves analogous to those plotted in Fig. \ref{fig17}. The second phase transition (suggesting hadronic stability at extremely high temperatures) corresponds to highly unstable confined states, due to their thermodynamic properties. Moreover, as the rotational velocity decreases (see Fig. \ref{fig18}), the second phase transition shifts rapidly toward $\bar{z}_h \rightarrow 0$, even at low densities (red line). In the non-relativistic limit, their curves exhibit behavior similar to that of first-order transitions at high densities, with the plasma being the dominant phase at high temperatures. (For non-rotating matter, this second phase transition disappears completely, see Fig. \ref{fig1}.) This way, the transitions for $\omega l \leq 0.16$ will not make a significant contribution to phase mixing in the intermediate regions between low and extremely high temperatures. In other words, the coexistence between the phases should be attributed solely to the $E \rightarrow F$ transitions, due to relativistic rotations in Region II.  

A similar analysis can be performed in high-density regimes based on the numerical results obtained in Ref. \cite{Junqueira:2025xnx}, involving the values of the maximum critical density at low temperatures --- see the critical $\omega_0(\mu)$ curve at zero temperature for the exact Andreev's model of Fig. 5 in this paper. The value of $\bar{\mu}$ at $T =0$ ($\bar{\mu}_0$) increases as the rotational velocity increases, until it reaches the value of $\bar{\mu}_0$ for the non-rotating system. The effect of rotation becomes more pronounced for $\omega l$ greater than $10\%$ of the speed of light; it shows a sharp increase up to $\omega l \approx 0.90$. For values less than $10\%$  the speed of light, the values of $\bar{\mu}_0(\omega l = 0)$ and $\bar{\mu}_0(\omega l)$ become practically indistinguishable. This demonstrates that the coexistence of phases for high-energy states should not appear dominant in the phase diagram, since in this region the temperatures are lower, and such mixing could only occur for $T < T_c (\omega l = 0)$, corresponding to systems that probably do not have very high rotational velocities. Above $T_c (\omega l = 0)$, the matter is always in the plasma phase, independently of its angular momentum. Below it, coexistence would occur only in a narrow region where the effect would be negligible. In this case, the transitions will be dominated by first-order transitions defined by the $T_c$ curve for non-rotating matter. The relativistic effect of rotation on the QCD phase diagram should be strong due to $E \rightarrow F$ transitions, which occur in the low-energy regime at high temperatures. 

\section{Estimation for the critical point of the QCD phase diagram}
\label{sec7}

From this analysis, we can make a holographic prediction of the critical point (CP) between smooth (crossover) transitions at low densities and first-order transitions in the high-density regime dominated by the $T_c(\omega l = 0)$ curve. The smooth transitions are a consequence of $E \rightarrow F$ transitions that allow the coexistence of confined and deconfined phases, which occur in the interval 
\begin{equation}\label{interval}
0 < \bar{\mu} \leq \text{max}(\bar{\mu}_{EF})\;,
\end{equation}
so that,
\begin{eqnarray}\label{muCP}
    \bar{\mu}_{CP} = \text{max}(\bar{\mu}_{EF}) \approx 0.414\;.   
\end{eqnarray}
The IR parameter $\sqrt{c}$ can be fixed in the soft wall model using QCD phenomenology. The fit of the masses of lightest $\rho$-mesons leads to $\sqrt{c}=338\, \text{MeV}$\footnote{This phenomenological estimate of $\sqrt{c}$ was obtained using a zero-density holographic soft wall model with $A_z = 0$, see \cite{Herzog:2006ra}. We do not rule out a possible refinement of this value, assuming the effect of rotation (where $A_z \neq 0$) and of the quark chemical potential. } \cite{Herzog:2006ra}. The $\eta$ parameter, which appears in the relation between the BH charge parameter and the quark chemical potential, see \eqref{qmu}, affects the HP transitions in a non-trivial way \cite{Colangelo:2010pe}. For simplicity, we have assumed $\eta = 1$. To recover its dependence, one must define the physical quark density  
\begin{eqnarray}
    \bar{\mu}_{\textsc{phys}}(\eta) = \eta \bar{\mu}\;. 
\end{eqnarray}
For a QCD system with $N_c$ colors and $N_f$ flavors, $\eta$ is given by the following expression in holographic models    \cite{Lee:2009bya}:
\begin{equation}\label{eta}
    \eta  = \sqrt{\frac{3N_c}{2N_f}}\;.
\end{equation}
By taking $N_c = 3$ and $N_f = 6$, from the characteristic color and flavor numbers of QCD, one finds $\eta = \sqrt{3/4}$. Using Eq. \eqref{varsw} and the phenomenological value of $\sqrt{c}$, the physical Andreev's estimate for the quark chemical potential at the critical point is given by 
\begin{eqnarray}
    {\mu_{CP}}_{\textsc{phys}} = \eta \sqrt{c}\bar{\mu}_{CP}  \approx \, 121.185\, \text{MeV} \;. 
\end{eqnarray}
The baryon chemical potentials are given by $\mu_B  = 3 \mu$, see \cite{Colangelo:2010pe}, so that the holographic prediction for the baryon density at CP for $N_f = 6$ is
\begin{eqnarray}\label{muB}
    \mu_{CPB}(N_f = 6) \approx 363.554 \, \text{MeV}\;. 
\end{eqnarray}
This estimation is strongly affected by the number of flavors, since $\eta$ is inversely proportional to $\sqrt{N_f}$, see Eq. \eqref{eta}. For $N_f = 3$, one finds
\begin{eqnarray}\label{muB2}
    \mu_{CPB}(N_f = 3) \approx 514.143 \, \text{MeV}\;. 
\end{eqnarray}
The chemical potential measured by an observer at a given frame tends to decrease as the rotational velocity increases. The estimations above are valid for an observer in a non-rotating frame. For the rotating one, we must use Eq. \eqref{muprime} with $\omega l \approx 0.3$ at CP, which yields
\begin{eqnarray}
 \mu_{CPB}^\prime(N_f = 6) &\approx& 346.808 \, \text{MeV}\;, \label{muBprime1}\\ 
    \mu_{CPB}^\prime(N_f = 3) &\approx& 460.461 \label{muBprime2}\, \text{MeV}\;. 
\end{eqnarray}
The relation between the chemical potential in a rotating frame and that in a frame at rest is given by Eq. \eqref{muprime}. The temperature at CP can be estimated using Eq. \eqref{barT}. The maximum value of $\bar{\mu}_{EF}$ occurs for a matter with rotational velocity $\omega l  \approx 0.3$, with critical horizon $\bar{z}_{h_c} \approx 0.832166$. Replacing these values into \eqref{barT}, for $\bar{\mu} = \bar{\mu}_{CP}$ given by \eqref{muCP}, one finds
\begin{equation}
    T_{CP} \approx 58.307 \, \text{MeV}\;.
\end{equation}
The point $(\mu_{CP}, T_{CP})$ defines the critical point uniquely, between the smooth transitions at low densities and first-order ones for $\mu \geq \mu_{CP}$, dominated by $T_c(\omega l\approx 0)$. The most critical baryon density at $T = 0$ in the exact Andreev model is estimated to be $\bar{\mu}_{0} \approx 1.067$ for $\eta =1$. Thus,  
\begin{eqnarray}
 \mu_{0B}(N_f = 6)  &\approx&   937.425 \, \text{MeV}\;, \label{mu01} \\
 \mu_{0B}(N_f = 3)   &\approx& 1325.100 \, \text{MeV}\;,\label{mu02}
\end{eqnarray}
such that $\mu_{CPB}/\mu_{0B} \approx 38.8\%$, which shows that the smooth transitions must occur before the first half of the QCD phase diagram. The values of \eqref{mu01} and \eqref{mu02} are also valid for the rotating frame; this transition is dominated by the non-rotating $T_c$ curve with $\omega l \approx 0$. On the other hand, the results \eqref{muBprime1} and \eqref{muBprime2} are sensitive even to small changes in the rotational velocity, which were estimated numerically at the critical point, with an accuracy of $\omega l = 0.30 \pm \delta \omega$, with $\delta \omega = 0.01$.

\section{Comparison with lattice QCD, effective models and holography}
\label{sec8}
The phase structure obtained in this work can be compared with results from lattice QCD, effective models, and previous holographic approaches, including rotation. Lattice simulations at finite angular velocity indicate that rotation tends to reduce the critical temperature of deconfinement, with the shift depending on the magnitude of the vorticity and the system size \cite{Chen:2015gta,Braguta:2021jgn,Chernodub:2016kxh}. This qualitative behavior is consistent with Eq. \eqref{HTrot}, where the factor $\sqrt{1-\omega^2 l^2}$ suppresses the temperature, leading to a decrease of $T_c$ with increasing rotation. Our results reproduce this tendency both at zero and finite density, and extend it by showing that, in the relativistic regime, rotation also modifies the order of the transition through the emergence of phase coexistence regions associated with $E \rightarrow F$ transitions.

In effective approaches such as NJL and PNJL models, rotation is known to affect chiral symmetry restoration and deconfinement, typically lowering the critical temperature and shifting the critical endpoint toward smaller chemical potentials \cite{Jiang:2016wvv,Fukushima:2003fw,Ratti:2005jh}. The mechanism is usually attributed to rotational suppression of condensates and modification of the fermionic spectrum. Our holographic results are compatible with this trend: the decrease of $T_c$ with $\omega l$ and the existence of a finite critical chemical potential separating crossover and first-order regions are qualitatively similar. However, in contrast to NJL/PNJL models, the present approach provides a geometric interpretation in terms of competing bulk solutions. Naturally, it incorporates strong-coupling effects, thereby identifying a mixed phase driven by angular-momentum distributions rather than solely by chiral dynamics.

The predicted critical point can also be compared quantitatively with other approaches. In the present model, we obtain $(\mu_B, T)_{CP} \approx (363\,\text{MeV} $--$ 514\,\text{MeV}, 58\,\text{MeV})$. Lattice extrapolations typically suggest a critical point, if present, in the range $\mu_B \sim 300$ - $600\,\text{MeV}$ and $T \sim 100$ - $160\,\text{MeV}$, although large uncertainties remain \cite{Borsanyi:2020fev,Stephanov:2004wx}. NJL and PNJL models often predict $T_{CP}$ between $50$ and $120\,\text{MeV}$ and $\mu_B$ between $300$ and $800\,\text{MeV}$, depending on parameter choices \cite{Fukushima:2003fw,Ratti:2005jh}. Compared to these results, our prediction lies in the lower-temperature region but within the expected range of baryon chemical potentials. The relatively low value of $T_{CP}$ can be attributed to the soft-wall scale and to the strong suppression induced by relativistic rotation, which effectively shifts the crossover region toward smaller temperatures.

\section{Conclusions}
\label{sec9}

In this work, we have studied the phase structure of rotating QCD matter within Andreev's soft-wall holographic model, considering a charged BH with nonzero angular momentum in AdS$_5$ spacetime. The Hawking-Page transitions were analyzed to describe the deconfinement of hadronic matter into a QGP at finite density and rotation.  
Our results reveal a rich interplay between chemical potential, temperature, and angular velocity. For quark chemical potentials above the maximum $\bar{\mu}_{EF}$, the phase transitions are predominantly of first order. The critical temperature decreases with increasing chemical potential, and the transitions remain sharp, reflecting conventional confinement-deconfinement behavior. In the non-relativistic limit, corresponding to small rotational velocities ($\omega l \lesssim 0.16$), the transitions are well-described by the critical temperature $T_c(\omega l \approx 0)$, and the coexistence of phases is negligible, even at intermediate densities. The second phase transition that may occur at extremely high temperatures corresponds to highly unstable hadronic states and is disregarded, due to the vanishing QCD coupling constant and their thermodynamic properties.  

On the other hand, for chemical potentials below $\text{max}(\bar{\mu}_{EF}) \approx 0.414$, relativistic rotations play a significant role in shaping the phase diagram. Assuming an ensemble with $N_p$ particles, being $N_p$ large, at high temperatures and low densities, where kinetic energy is abundant, the QGP and hadronic matter coexist over a wide range of angular momenta. In this region, transitions occur smoothly rather than abruptly, resulting in crossover-like behavior rather than first-order transitions. This interpretation can be explained from the analysis of the thermodynamic properties of the system (Gibbs free energy, entropy, specific heat) for each phase, which was carried out in Section 6, using holographic renormalization. The negative QCD coupling constant influences the coexistence region, and the total Gibbs free energy must be considered as an infinite sum over all coexisting states. These smooth transitions occur predominantly in intermediate-low-density regions, whereas at lower temperatures, hadronic matter remains stable and the transitions are governed by the non-rotating critical temperature.  Our analysis demonstrates that the relativistic effect of rotation is most pronounced for transitions labeled $E \rightarrow F$, occurring at high temperatures and low densities. At lower rotational velocities or in high-density, low-temperature regimes, the coexistence of phases is negligible, and first-order transitions dominate. Overall, the exact soft-wall model provides a robust framework to capture both conventional first-order transitions at high densities and rotationally induced smooth crossover transitions at low densities, highlighting the significant impact of angular momentum on the QCD phase diagram.

\begin{figure}[!htb]
	\centering
    \includegraphics[scale=0.55]{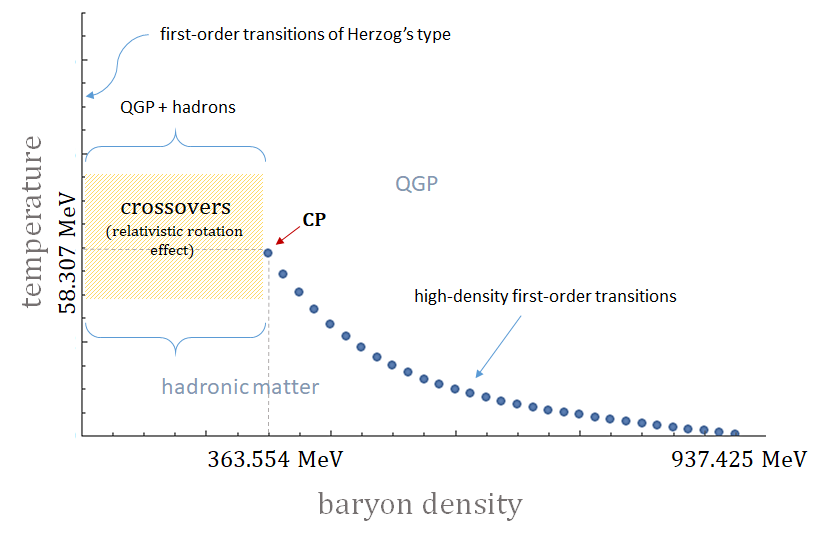}
	\caption{QCD phase diagram in the exact Andreev's holographic soft wall model.}
    \label{fig19}
\end{figure}

In Fig. \ref{fig19}, we plot the QCD phase diagram including all results from the analysis of phase transitions at low and high densities of strongly interacting matter in Andreev's exact soft-wall AdS/QCD model, accounting for relativistic rotations. Between the first-order Herzog-type transition at zero density and those occurring in the high-density regime, there is an intermediate region in which hadronic matter and plasma can coexist with different angular momenta. Crossover transitions characterize this intermediate region, since it is described by $E \rightarrow F$ transitions, see Sec. \ref{sec6}, whose states vary smoothly in the phase diagram, rather than by jumping from one phase to another. This behavior is observed in the low-density regime of rotating matter, as shown in Figs.  \ref{fig5} and \ref{fig6}, in which one can observe the existence of an interval where the HP transition no longer occurs, and the system is always in the confined phase. The critical chemical potential that defines this interval in which hadrons are always stable depends on the rotational velocity, see Fig. \ref{fig7}. For this reason, the holographic description indicates the existence of a mixed phase that can occur at low densities and high temperatures, as inferred from the analysis in Fig. \ref{fig8}. The subdivisions shown in this figure could be made infinitely, corresponding to smooth point-to-point transitions (crossovers), according to the phases mixing defined by the angular momentum of the particles.

The interval over which crossover transitions occur is defined by Eq. \eqref{interval}, as can be inferred from the analysis of Fig. \ref{fig7}. There is a maximum value of the chemical potential for which this type of coexistence between phases can occur for states with distinct angular momentum. This defines the critical point between smooth transitions at low densities and first-order transitions at high densities and lower temperatures. In terms of baryonic density, the prediction of Andreev's holographic model in the non-rotating frame, for a system with $N_f=6$ flavors, is given by Eqs. (\ref{muB}, \ref{muBprime2}). 
These values were obtained through numerical and phenomenological analysis described in Sec. \ref{sec7}. The CP for $N_f = 6$ is indicated in Fig. \ref{fig17}. For $\mu \geq \mu_{CPB} $, the system is described by first-order transitions, dominated by the $T_c(\omega l \approx 0)$ curve, for states with rotational velocities much lower than the speed of light, as discussed in Sec. \ref{sec6.2}. The distinction between the types of phase transitions at high and low densities is attributed to the effect of relativistic rotations, which can only occur for $\omega l \gtrsim 0.16$, in the low-density regime ($\mu \leq \mu_{CPB})$, see Fig. \ref{fig7}. This estimate indicates that this effect should be considerable in states at very high temperatures, as shown by the yellow-hatched region in Fig. \ref{fig19}. In the non-relativistic limit, the $E \to F$  transitions cease to occur and are automatically described by first-order transitions, in which hadron matter is more stable at lower temperatures, see Fig. \ref{fig18}. Meanwhile, at the limit of extremely high temperatures, it is expected that the hadronic matter will become unstable, due to the negative $\beta$-function of QCD. This instability is consistently described by the present holographic model, since the specific heat of hadronic states (dual to the thermal AdS geometry) becomes negative as the temperature increases, see Fig. \ref{fig13}, indicating a marked loss of energy of these states. In this case, the transition will occur smoothly, from a mixed state -- QGP and hadrons with different angular momentum -- to a phase dominated by QGP states at extremely high temperatures. On the other hand, at the low-temperature limit, hadronic matter is dominant. These transitions, from stable hadronic matter at lower temperatures to a mixed phase, and then from the mixed phase to a phase energetically dominated by QGP states, are consistently described by the total energy of the system in the grand-canonical ensemble, see Fig. \ref{fig16}. Such transitions are indeed smooth, as demonstrated by Figures \ref{fig11} and \ref{fig12}, in which the entropy and specific heat curves for the ensemble do not indicate any discontinuities.

Building on the comprehensive analysis presented here of rotational effects on the QCD phase diagram within an exact soft-wall holographic model, several concrete perspectives emerge to further refine the understanding of rotating strongly interacting matter. A natural next step is to incorporate the full backreaction of the rotating charged geometry in AdS, enabling an even more precise determination of the interface between crossover behavior and first-order transitions. Incorporating subleading dilaton and higher-derivative corrections could test the stability of the coexistence window by revealing whether the smooth transition region persists once the AdS bulk dynamics is modified beyond the leading soft-wall approximation, or whether it collapses into a sharper first-order structure when higher-order contributions to the action and metric response are taken into account. Moreover, calculating two-point functions and quasinormal spectra for AdS bulk perturbations dual to both the hadronic phase and the QGP plasma states with different boundary angular momenta would clarify whether the mixed configurations identified here manifest distinct dynamical signatures. The specific heat, which quantifies how the system’s energy responds to temperature changes, and the baryon number susceptibility, which measures how baryon density responds to chemical potential, both serve as probes of the phase structure. Peaks or divergences in these quantities signal the approach to a critical point and can be used to locate the transition between crossover and first-order behavior independently. Together, these steps would refine the predictive power of the soft-wall model in the QGP rotational regime.

\subsection*{Acknowledgements}
OCJ thanks The S\~ao Paulo Research Foundation (FAPESP) 
(Grants No. 2021/01089-1 and No. 2024/14390-0). The work of RdR is supported by FAPESP  (Grants No. 2021/01089-1 and No. 2024/05676-7) and the National Council for Scientific and Technological Development - CNPq  (Grants No. 303742/2023-2 and No. 401567/2023-0).

\appendix

\section{Auxiliary tables with numerical results} \label{app1}

The values of the variables in the tables are in their dimensionless versions. To obtain these values in MeV units, simply apply equations \eqref{varsw} and \eqref{barT}, with $\sqrt{c} = 338\, \text{MeV}$.  

\begin{table}[h]
\centering
\begin{tabular}[c]{|c|!{\vrule width 1pt}c|c|}
\hline 
\text{}   $\,\,\,\bar{\mu}\,\,\, $ & $\,\,\,\bar{z}_{hc}\,\,\, $& $\,\,\,\bar{T}_c\,\,\, $\\
 \hline\hline
 $\,\,\,0.00\,\,\, $ &$\,\,\,0.647329\,\,\,$ &$\,\,\,0.907534\,\,\,$ \\
\hline
  $\,\,\,0.05\,\,\, $ &$\,\,\,0.650309\,\,\,$&$\,\,\,0.883106\,\,\,$   \\
\hline
  $\,\,\,0.10\,\,\, $ &$\,\,\,0.659150\,\,\,$  &$\,\,\,0.814762\,\,\,$ \\
\hline
 $\,\,\,0.15\,\,\, $ &$\,\,\,0.673549\,\,\,$&$\,\,\,0.715485\,\,\,$  \\
\hline
  $\,\,\,0.20\,\,\, $ &$\,\,\,0.692962\,\,\,$ &$\,\,\,0.601822\,\,\,$ \\
\hline
 $\,\,\,0.25\,\,\, $ &$\,\,\,0.716562\,\,\,$&$\,\,\,0.489041\,\,\,$  \\
\hline
 $\,\,\,0.30\,\,\, $ &$\,\,\,0.743208\,\,\,$ &$\,\,\,0.387911\,\,\,$ \\
\hline
 $\,\,\,0.35\,\,\, $ &$\,\,\,0.771485\,\,\,$&$\,\,\,0.303772\,\,\,$  \\
\hline
 $\,\,\,0.40\,\,\, $ &$\,\,\,0.799867\,\,\,$&$\,\,\, 0.237423\,\,\,$  \\
\hline
 $\,\,\,0.45\,\,\, $ &$\,\,\,0.826967\,\,\,$&$\,\,\,0.186826\,\,\,$  \\
\hline
 $\,\,\,0.50\,\,\, $ &$\,\,\,0.851777\,\,\,$&$\,\,\, 0.148783\,\,\,$  \\
\hline
 $\,\,\,0.55\,\,\, $ &$\,\,\,0.873768\,\,\,$&$\,\,\,0.120083\,\,\,$  \\
\hline
 $\,\,\,0.60\,\,\, $ &$\,\,\,0.892832\,\,\,$&$\,\,\,0.0980534\,\,\,$  \\
\hline
  $\,\,\,0.65\,\,\, $ &$\,\,\,0.909137\,\,\,$ &$\,\,\,0.0806795\,\,\,$ \\
\hline
 $\,\,\,0.70\,\, $ &$\,\,\,0.922993\,\,\,$&$\,\,\,0.0665263\,\,\,$   \\
\hline
  $\,\,\,0.75\,\,\, $ &$\,\,\,0.934746\,\,\,$&$\,\,\,0.0545959\,\,\,$   \\
\hline
 $\,\,\,0.80\,\,\, $ &$\,\,\,0.944729\,\,\,$&$\,\,\,0.0441995\,\,\,$  \\
\hline
  $\,\,\,0.85\,\,\, $ &$\,\,\,0.953237\,\,\,$ &$\,\,\,0.0348597\,\,\,$ \\
\hline
 $\,\,\,0.90\,\, $ &$\,\,\,0.960521\,\,\,$&$\,\,\,0.0262422\,\,\,$   \\
\hline
  $\,\,\,0.95\,\,\, $ &$\,\,\,0.966787\,\,\,$&$\,\,\,0.0181095\,\,\,$   \\
\hline
 $\,\,\,1.00\,\, $ &$\,\,\,0.972207\,\,\,$&$\,\,\,0.0102904\,\,\,$   \\
\hline
  $\,\,\,1.05\,\,\, $ &$\,\,\,0.976919\,\,\,$&$\,\,\,0.00265945\,\,\,$   \\
\hline
 \end{tabular}   
\caption{Quark chemical potentials, critical horizons and critical temperatures of deconfinement for non-rotating matter, used to plot Fig. \ref{fig2}. }
\label{table1}
\end{table} 

\begin{table}[h]
\centering
\begin{tabular}[c]{|c|!{\vrule width 1pt}c|}
\hline 
\text{}   $\,\,\,\bar{\mu}_{EF}\,\,\, $ & $\,\,\,\omega l  \,\,\, $\\
 \hline\hline
 $\,\,\,0.340\,\,\, $ &$\,\,\,0.16\,\,\,$ \\
\hline
  $\,\,\,0.375\,\,\, $ &$\,\,\,0.18\,\,\,$  \\
\hline
  $\,\,\, 0.393\,\,\, $ &$\,\,\,0.20\,\,\,$   \\
\hline
 $\,\,\,0.414 \,\,\, $ &$\,\,\,0.30\,\,\,$  \\
\hline
  $\,\,\,0.396\,\,\, $ &$\,\,\,0.40\,\,\,$ \\
\hline
 $\,\,\,0.362\,\,\, $ &$\,\,\,0.50\,\,\,$ \\
\hline
 $\,\,\, 0.320\,\,\, $ &$\,\,\,0.60\,\,\,$  \\
\hline
 $\,\,\, 0.270\,\,\, $ &$\,\,\,0.70\,\,\,$  \\
\hline
 $\,\,\,0.209\,\,\, $ &$\,\,\,0.80\,\,\,$  \\
\hline
 $\,\,\,0.132\,\,\, $ &$\,\,\,0.90\,\,\,$  \\
\hline
 $\,\,\, 0.0767\,\,\, $ &$\,\,\,0.95\,\,\,$  \\
\hline
 $\,\,\, 0.0169\,\,\, $ &$\,\,\,0.99\,\,\,$  \\
\hline
 \end{tabular}   
\caption{Critical values of $\bar{\mu}_{EF}$ at different rotational velocities, used to plot Fig. \ref{fig7}.}
\label{table2}
\end{table} 

\begin{table}[h]
\centering
\begin{tabular}[c]{|c|!{\vrule width 1pt}c|c|}
\hline 
\text{}   $\,\,\,\bar{\mu}\,\,\, $ & $\,\,\,\bar{z}_{hc}\,\,\, $& $\,\,\,\bar{T}_c\,\,\, $\\
 \hline\hline
  $\,\,\,0.414\,\,\, $ &$\,\,\,0.832166\,\,\,$ &$\,\,\,0.172505\,\,\,$ \\
\hline
 $\,\,\,0.417\,\,\, $ &$\,\,\,0.891007\,\,\,$ &$\,\,\,0.105907\,\,\,$ \\
\hline
  $\,\,\,0.420\,\,\, $ &$\,\,\,0.916365\,\,\,$&$\,\,\,0.0857094\,\,\,$   \\
\hline
  $\,\,\,0.425\,\,\, $ &$\,\,\,0.947759\,\,\,$  &$\,\,\,0.0657364\,\,\,$ \\
\hline
 $\,\,\,0.430\,\,\, $ &$\,\,\,0.973016\,\,\,$&$\,\,\,0.0528738\,\,\,$  \\
\hline
  $\,\,\,0.435\,\,\, $ &$\,\,\,0.994843\,\,\,$ &$\,\,\,0.0436076\,\,\,$ \\
\hline
 $\,\,\,0.440\,\,\, $ &$\,\,\,1.01437\,\,\,$&$\,\,\,0.0365297\,\,\,$  \\
\hline
 $\,\,\,0.445\,\,\, $ &$\,\,\,1.03219\,\,\,$ &$\,\,\,0.0309204\,\,\,$ \\
\hline
 $\,\,\,0.450\,\,\, $ &$\,\,\,1.04867\,\,\,$&$\,\,\,0.0263589\,\,\,$  \\
\hline
 $\,\,\,0.455\,\,\, $ &$\,\,\,1.06406\,\,\,$&$\,\,\, 0.0225771\,\,\,$  \\
\hline
 $\,\,\,0.460\,\,\, $ &$\,\,\,1.07853\,\,\,$&$\,\,\,0.019394\,\,\,$  \\
\hline
 $\,\,\,0.465\,\,\, $ &$\,\,\,1.09220\,\,\,$&$\,\,\, 0.0166815\,\,\,$  \\
\hline
 $\,\,\,0.470\,\,\, $ &$\,\,\,1.10517\,\,\,$&$\,\,\,0.0143462\,\,\,$  \\
\hline
 $\,\,\,0.475\,\,\, $ &$\,\,\,1.11753\,\,\,$&$\,\,\,0.0123177\,\,\,$  \\
\hline
  $\,\,\,0.480\,\,\, $ &$\,\,\,1.12933\,\,\,$ &$\,\,\,0.0105423\,\,\,$ \\
\hline
 $\,\,\,0.490\,\, $ &$\,\,\,1.15148\,\,\,$&$\,\,\,0.00759059\,\,\,$   \\
\hline
  $\,\,\,0.500\,\,\, $ &$\,\,\,1.17196\,\,\,$&$\,\,\,0.00524644\,\,\,$   \\
\hline
 $\,\,\,0.510\,\,\, $ &$\,\,\,1.19103\,\,\,$&$\,\,\,0.00334881\,\,\,$  \\
\hline
  $\,\,\,0.520\,\,\, $ &$\,\,\,1.20887\,\,\,$ &$\,\,\,0.00178761\,\,\,$ \\
\hline
 $\,\,\,0.530\,\, $ &$\,\,\,1.22565\,\,\,$&$\,\,\,0.000485152\,\,\,$   \\
\hline
 \end{tabular}   
\caption{Quark chemical potentials, critical horizons and critical temperatures of deconfinement at $\omega l = 0.3$, used in Fig. \ref{fig8}. }
\label{table3}
\end{table} 

\begin{table}[h]
\centering
\begin{tabular}[c]{|c|!{\vrule width 1pt}c|c|}
\hline 
\text{}   $\,\,\,\bar{\mu}\,\,\, $ & $\,\,\,\bar{z}_{hc}\,\,\, $& $\,\,\,\bar{T}_c\,\,\, $\\
 \hline\hline
  $\,\,\,0.3960\,\,\, $ &$\,\,\,0.986429\,\,\,$ &$\,\,\,0.0466958\,\,\,$ \\
\hline
 $\,\,\,0.3962\,\,\, $ &$\,\,\,0.997143\,\,\,$ &$\,\,\,0.0426659\,\,\,$ \\
\hline
  $\,\,\,0.3965\,\,\, $ &$\,\,\,1.00682\,\,\,$&$\,\,\,0.0393038\,\,\,$   \\
\hline
  $\,\,\,0.3970\,\,\, $ &$\,\,\,1.01827\,\,\,$  &$\,\,\,0.0356338\,\,\,$ \\
\hline
  $\,\,\,0.3990\,\,\, $ &$\,\,\,1.04778\,\,\,$ &$\,\,\,0.0275359\,\,\,$ \\
\hline
 $\,\,\,0.4025\,\,\, $ &$\,\,\,1.08134\,\,\,$ &$\,\,\,0.0202862\,\,\,$ \\
\hline
 $\,\,\,0.4050\,\,\, $ &$\,\,\,1.09996\,\,\,$&$\,\,\,0.0169900\,\,\,$  \\
\hline
 $\,\,\,0.4100\,\,\, $ &$\,\,\,1.13068\,\,\,$&$\,\,\, 0.0124675\,\,\,$  \\
\hline
 $\,\,\,0.4150\,\,\, $ &$\,\,\,1.15619\,\,\,$&$\,\,\,0.00942661\,\,\,$  \\
\hline
 $\,\,\,0.4200\,\,\, $ &$\,\,\,1.17840\,\,\,$&$\,\,\, 0.00721243\,\,\,$  \\
\hline
 $\,\,\,0.4300\,\,\, $ &$\,\,\,1.21633\,\,\,$&$\,\,\,0.00418558\,\,\,$  \\
\hline
 $\,\,\,0.4400\,\,\, $ &$\,\,\,1.24844\,\,\,$&$\,\,\,0.00221573\,\,\,$  \\
\hline
  $\,\,\,0.4500\,\,\, $ &$\,\,\,1.27653\,\,\,$ &$\,\,\,0.000840589\,\,\,$ \\
\hline
 \end{tabular}   
\caption{Quark chemical potentials, critical horizons and critical temperatures of deconfinement at $\omega l = 0.4$, used in Fig. \ref{fig8}. }
\label{table4}
\end{table} 

\begin{table}[h]
\centering
\begin{tabular}[c]{|c|!{\vrule width 1pt}c|c|}
\hline 
\text{}   $\,\,\,\bar{\mu}\,\,\, $ & $\,\,\,\bar{z}_{hc}\,\,\, $& $\,\,\,\bar{T}_c\,\,\, $\\
 \hline\hline
  $\,\,\,0.362\,\,\, $ &$\,\,\,1.12287\,\,\,$ &$\,\,\,0.0141584\,\,\,$ \\
\hline
 $\,\,\,0.365\,\,\, $ &$\,\,\,1.18679\,\,\,$ &$\,\,\,0.00763720\,\,\,$ \\
\hline
  $\,\,\,0.370\,\,\, $ &$\,\,\,1.23857\,\,\,$&$\,\,\,0.00425500\,\,\,$   \\
\hline
  $\,\,\,0.375\,\,\, $ &$\,\,\,1.27333\,\,\,$  &$\,\,\,0.00263073\,\,\,$ \\
\hline
  $\,\,\,0.380\,\,\, $ &$\,\,\,1.30113\,\,\,$ &$\,\,\,0.00161183\,\,\,$ \\
\hline
 $\,\,\,0.385\,\,\, $ &$\,\,\,1.32483\,\,\,$ &$\,\,\,0.000902190\,\,\,$ \\
\hline
 $\,\,\,0.390\,\,\, $ &$\,\,\,1.34573\,\,\,$&$\,\,\,0.000377878\,\,\,$  \\
\hline
 \end{tabular}   
\caption{Quark chemical potentials, critical horizons and critical temperatures of deconfinement at $\omega l = 0.5$, used in Fig. \ref{fig8}. }
\label{table5}
\end{table}

\clearpage

\bibliographystyle{utphys2}
\bibliography{library}

\end{document}